\documentclass[aps,prb,twocolumn,superscriptaddress]{revtex4-2}

\usepackage{amsmath}
\usepackage{amsfonts}
\usepackage{amssymb}
\usepackage[colorlinks=true,citecolor=blue,linkcolor=red,pdfduplex=DuplexFlipLongEdge]{hyperref}
\usepackage{graphicx}
\usepackage[dvipsnames]{xcolor}
\usepackage{newtxtext}

\usepackage{bm} 
\usepackage{dcolumn}
\usepackage{mathtools}

\usepackage{array,booktabs,ragged2e}	
\usepackage{soul}						%
\usepackage{cancel}						%
\usepackage{tabularx}%
\usepackage{multirow}
\usepackage{hhline}
\usepackage{adjustbox}
\usepackage{sidecap}
\usepackage{threeparttable}
\usepackage{colortbl} %
\usepackage{tabularx}
\usepackage[utf8]{inputenc}
\usepackage[T1]{fontenc}
\usepackage{quiver}

\newcolumntype{P}[1]{>{\raggedleft\arraybackslash}p{#1}}
\newcolumntype{R}[1]{>{\centering\arraybackslash}p{#1}}

\usepackage{bm,color,amsmath,amssymb,mathrsfs,latexsym,graphicx,psfrag,mathtools}
\usepackage{color}
\usepackage[dvipsnames]{xcolor}

\usepackage{empheq}%

\newcommand{\bsl}[1]{\boldsymbol{#1}}

\newcommand{\ii}{\mathrm{i}}

\newcommand{\appref}[1]{Appendix.\ref{#1}}
\newcommand{\refcite}[1]{Ref.\cite{#1}}

\newcommand{\eq}[1]{\begin{equation} #1 \end{equation}}

\newcommand{\eqa}[1]{\begin{align}\begin{split} #1 \end{split}\end{align}}

\let\oldAA\AA
\renewcommand{\AA}{\text{\normalfont\oldAA}}

\newcommand{\ie}{{\emph{i.e.}}}
\newcommand{\eg}{{\emph{e.g.}}}

\newcommand{\A}{\mathcal{A}}

\newcommand{\I}{\mathcal{I}}

\usepackage{comment}

\usepackage{cleveref}
\crefname{appendix}{App.}{Apps.}
\crefname{equation}{Eq.}{Eqs.}
\crefname{figure}{Fig.}{Figs.}
\crefname{table}{Tab.}{Tabs.}
\crefname{section}{Sec.}{Secs.}
\creflabelformat{appendix}{[#2#1#3]}

\usepackage{amsmath,amssymb,amsthm}
\theoremstyle{definition}

\begin{document}

\title{Nematic Wigner Crystals in Rhombohedral Multilayer Graphene}

\author{Yang Ge}
\author{Ashwini Malviya}
\author{Ethan Angerhofer}
\affiliation{Department of Physics and Quantum Theory Project, University of Florida, Gainesville, FL 32611, USA}
\author{Zhengguang Lu}
\affiliation{Physics Department, Florida State University, Tallahassee, FL 32306, USA}
\author{Jiabin Yu}
\email{yujiabin@ufl.edu}
\affiliation{Department of Physics and Quantum Theory Project, University of Florida, Gainesville, FL 32611, USA}

\begin{abstract}

Recent experiments have reported evidence for Wigner crystals (WCs) in rhombohedral graphene.
Here, we investigate Wigner crystallization in rhombohedral tetralayer graphene using projected Hartree--Fock (HF) calculations and time--dependent Hartree--Fock (TDHF) calculations.
We first perform HF calculations with one electron per Wigner unit cell, and find nematic WCs (nWCs) that spontaneously break the threefold rotational symmetry $C_3$ and $C_3$-invariant WCs.
In particular, there are two nWC regions in the phase diagram: one larger region at large displacement fields and low electron densities, and another smaller region at intermediate fields and high densities.
Both the nWCs and the $C_3$-invariant WCs are valley-polarized states with zero Chern number, and have positive indirect gaps in the HF band structure.
We then perform TDHF calculations to further test the local stability of the WC states.
We find that all $C_3$-invariant WCs and half of the nWCs are locally stable, while the remaining nWCs are unstable towards WCs with two electrons per unit cell or metallic states.
The predicted stable nWC phase can be identified experimentally by scanning tunneling microscopy through its anisotropic charge distribution or by angle-resolved transport measurements via a direction-dependent depinning voltage.

\end{abstract}

\maketitle

\section{Introduction}
Since Wigner's original proposal of electron crystallization driven by Coulomb repulsion \cite{Wigner_1934_WC1,Wigner_1938_WC2}, Wigner crystals (WCs) have served as a paradigmatic example of interaction-induced crystalline ordering that spontaneously breaks (approximate) continuous translation symmetry~\cite{Shayegan_2022_WC_review}.
Here, ``approximate'' means that the  unit cell of the WC must be much larger than that of the underlying material.
First realized with electrons floating above liquid ${}^4$He \cite{Grimes_1979_WignerHelium,Monarkha_2004_2DCoulomb} (see also the recent work \cite{Schleusner_2026_Wigner_Nonlinear},) electron Wigner crystallization has since been experimentally reported in many two-dimensional condensed-matter platforms.
These include semiconductor heterostructures with \cite{Andrei_1988_Magnetic_Wigner,Goldman_1990_QH_Wigner,Jiang_1990_QH_Wigner,DasSarma_1997_QHE_Perspectives,Liu_2016_Anisotropic_WC,Hossain_2022_Anisotropic_Disordered_WS,Piot_2008_Quasi3D_WC} and without a magnetic field \cite{Yoon_1999_Wigner_GaAs,Falson_2022_ZeroField_WC_ZnO}
as well as monolayer and bilayer transition-metal dichalcogenides at low densities \cite{Smolenski_2021_Monolayer_WC,Zhou_2021_BilayerWigner_TMD_Hetero,Xiang_2024_Disordered_WC_Melting,Wang_2025_Wigner_Crystal_Polarons,Zhang_2025_Wigner_Polarons_Dynamics,Munyan_2024_ZeroField_WC_Cd3As2,Tsui_2024_Magnetic_WC_Bilayer}.
All the WCs realized in these experiments form a triangular lattice and preserve the threefold rotation symmetry.

Then, a particularly intriguing question is whether the  Wigner crystal can \emph{spontaneously} break threefold rotational symmetry and form a nematic Wigner crystal (nWC).
Such a nWC has not been observed---the observed rotational-symmetry-breaking in WCs all comes from explicit and external symmetry breaking such as effective mass anisotropy \cite{xiang2026anisotropicwigner,Hossain_2022_Anisotropic_Disordered_WS} or an in-plane magnetic field \cite{Liu_2016_Anisotropic_WC}.
Previous theoretical studies on nWCs~\cite{Ettouhami_2006_Nematic_QH,Berg_2012_Nematic_SOC_2DEG,zhu2016nematic} either did not explicitly establish the nWC as the ground state~\cite{Berg_2012_Nematic_SOC_2DEG}, required a magnetic field to explicitly break continuous translational symmetry along one direction~\cite{Ettouhami_2006_Nematic_QH}, or focused on bosonic systems~\cite{zhu2016nematic}.
Therefore, the prediction of an electronic nWC without a magnetic field remains an open problem.

The system we focus on in this work is rhombohedral $n$-layer graphene (R$n$G), which has emerged as a remarkably tunable platform for correlated electronic phases.
Its chiral stacking produces low-energy bands whose bandwidth and quantum geometry can be controlled by the layer number and displacement field.
This flexibility has enabled the experimental observation of several correlated phases in rhombohedral graphene, including correlated insulating and metallic phases~\cite{Kerelsky_2021_Moireless_R4G,Zhou_2021_R3G_HalfQuarter,liu2024spontaneous}, orbital multiferroicity~\cite{han2023orbital}, superconducting phases~\cite{zhou2021superconductivity,Han_2025_TRB_SC_RMG,Nguyen_2025_SC_RMG,Seo_2025_SC_in_RMG,Deng_2025_SC_Ferroelectric_R6G,Yang_2026_MagSC,Zheng_2026_MultipleSC_R7G,Kalantre_2026_Fermiology_R4G,Kumar_2026_SC_Dual_Surface_Carriers,Nguyen_2026_CDW_SC}, as well as integer and fractional Chern insulating states~\cite{han2024correlated,han2024large,sha2024observation,Li_2026_MoireRMG_Chern}.
These experiments have stimulated theoretical work on correlated and topological phases intrinsic to rhombohedral graphene, including interaction-driven anomalous Hall and other electronic crystals~\cite{dong2024anomalous,soejima2024anomalous,dong2024stability,bernevig2025berry,miao_2026_RMG_crystal,Desrochers_2026_RMG_FAHC}; fractional quantum anomalous Hall phases without an  moir\'e potential~\cite{zhou2024fractional,huang2024self,huang2025fractional,Desrochers_2026_RMG_FAHC}; the elastic and sliding dynamics of electronic crystals~\cite{desrochers2025elastic,zeng2025berry}; as well as non-{Abelian} Chern bands, {Majorana} crystals, and valley-domain-wall transport~\cite{uchida2025non,Yoon_2026_Majorana_Crystal_RMG,Phong_2026_Valley_Valves_Domain_Walls}.

A remarkable recent development is the experimental evidence for both insulating and metallic Wigner crystals in rhombohedral tetralayer \cite{Han_2026_Wigner_R4G}, pentalayer \cite{Zhou_2026_RMG_WC}, and hexalayer \cite{Nguyen_2026_CDW_SC} graphene, without hBN alignment.
Despite the extensive related theoretical studies in rhombohedral graphene~\cite{dong2024stability,dong2024anomalous,soejima2024anomalous,zhou2024fractional,Dong_2026_mWC_RMG,Zaletel_2026_mWC_RMG,miao_2026_RMG_crystal,Dong_2025_Phonons_Berry}, the possible nematicity in WCs remains unexplored.
At the same time, both theoretical studies~\cite{Huang_2023_metallic_magnetism_R3G,Martinez_2025_Nematic_HF_RMG,Banerjee_2026_resistive_anisotropies} and experiments~\cite{Morissette_2025_Nematicity_R6G,Qin_2025_Anisotropy_RMG} indicate that metallic phases in rhombohedral graphene can spontaneously break rotational symmetry.
It is therefore natural to ask whether the WC phases in rhombohedral graphene can develop nematicity (\ie, spontaneously broken rotational symmetry).

In this work, we perform Hartree--Fock (HF) and time-dependent HF (TDHF) calculations for WC phases in R4G at low electron densities $n_e\in[0.2,0.4]\times10^{12}\,\text{cm}^{-2}$ and displacement fields corresponding to interlayer energy differences $V_{\text{disp}}\in[40,100]\,\text{meV}$.
At one electron per Wigner unit cell, the HF calculations show that the insulating WC phase spontaneously breaks the spinless threefold rotational symmetry ($C_3$) in R4G at relatively high displacement fields and low densities, as well as in a smaller region at intermediate fields and higher densities.
In contrast to the $C_3$-invariant WC, the nWC does not form a triangular lattice (\ie, a hexagonal two-dimensional Bravais lattice).
Both the nWC and the $C_3$-invariant WC phases that we find have zero Chern number.
Using TDHF calculations, we confirm stability in half of the nWC region, while the remaining nWC solutions are unstable.
One third of the unstable cases calculated have the instability towards a WC state with two electrons per cell \cite{Moulopoulos_1992_Paired_Crystal,Moulopoulos_1993_Paired_Theory,Taut_1994_No_Paired,Zverevich_2026_Triplet_Wigner,MacDonald_2026_RMG_Guide}, while the others are unstable towards metallic WCs where the electron filling is incommensurate~\cite{Han_2026_Wigner_R4G,Nguyen_2026_CDW_SC,Zhou_2026_RMG_WC,Dong_2026_mWC_RMG,Zaletel_2026_mWC_RMG}, Fermi liquids, or stripe phases.
Interestingly, we find that nWCs do not always outcompete the $C_3$-invariant WCs in correlation energy.
For some nWCs at low density and large displacement field, the electrons mainly occupy one lobe of the three-lobed Fermi surface that appears, and outcompete the $C_3$-invariant WC as they have a lower kinetic energy.
Experimentally, the nWC may be identified by visualizing its density profile with STM or by measuring the orientational dependence of its depinning voltage in transport~\cite{Csathy_2007_astable_Wigner}.

\section{Model}
\label{sec:model}

We begin with the continuum model for R$n$G introduced in \refcite{herzog2024moire}, including the spin and valley degrees of freedom.
The electron creation operator is $c^\dagger_{\bsl{k},l,\sigma,\eta,s}$, where $\bsl{k}$ is the continuum momentum, $\eta=\pm K$ labels the valley, $l=0,\ldots,3$ labels the layer, $\sigma=A,B$ labels the sublattice, and $s=\uparrow,\downarrow$ labels the spin. 
In the $K$ valley, the single-particle matrix Hamiltonian per spin is
\eq{
H_{K} = 
\begin{pmatrix}
v_F \bsl{k} \cdot \boldsymbol{\sigma} & t^{\dagger}(\bsl{k}) & t'^{\dagger} & \\
t(\bsl{k}) & \ddots & \ddots & t'^{\dagger} \\
t' & \ddots & v_F \bsl{k} \cdot \boldsymbol{\sigma} & t^{\dagger}(\bsl{k}) \\
 & t' & t(\bsl{k}) & v_F \bsl{k} \cdot \boldsymbol{\sigma}
\end{pmatrix} + H_{\text{ISP}}+ H_D,
}
with
\eq{
t(\bsl{k}) = -\begin{pmatrix} v_4 k & -t_1 \\ v_3 \bar{k} & v_4 k \end{pmatrix}, \quad 
t' = \begin{pmatrix} 0 & 0 \\ t_2 & 0 \end{pmatrix},
}
where $k\equiv k_x+i k_y$.
The Hamiltonian in valley $-K$ is obtained by time reversal.
$[H_{\text{ISP}}]_{l\sigma,l'\sigma'}=V_{\text{ISP}}|l-\frac{n-1}{2}|\delta_{ll'}\delta_{\sigma\sigma'}$ is the inversion-symmetric potential, and $[H_D]_{l\sigma,l'\sigma'}=V_{\text{disp}}(l-\frac{n-1}{2})\delta_{ll'}\delta_{\sigma\sigma'}$ captures the displacement field perpendicular to the sample, where $V_{\text{disp}}$ denotes the energy difference between neighboring layers induced by the displacement field. The parameters of the single-particle model are adopted from \refcite{herzog2024moire},
which is in excellent agreement with the density functional theory in \appref{app:DFT}.
The electrons repel each other via the double-gated screened Coulomb interaction
\eq{
H_\text{int} = \frac{1}{2 \A}\sum_{\bsl{q}}V(\bsl{q})
:{\rho}_{\bsl{q}}{\rho}_{-\bsl{q}}:,
}
where $\A$ is the sample area, $:\!...\!:$ denotes the normal ordering, $\rho_{\bsl{q}}=\sum_{\bsl{k},l,\sigma,\eta,s}c^\dagger_{\bsl{k}+\bsl{q},l,\sigma,\eta,s}c_{\bsl{k},l,\sigma,\eta,s}$, and $
V(\bsl{q}) = \pi \xi^2 V_\xi \frac{\tanh(\xi |\bsl{q}|/2)}{\xi |\bsl{q}|/2}$, where $V_\xi = \frac{e^2}{4\pi\epsilon_r\epsilon_0\xi}$,
the relative dielectric constant $\epsilon_r=5$, and the gate-to-gate distance $\xi=50\,\text{nm}$.

For the density and displacement-field ranges that we consider, \ie, $n_e\in[0.2,0.4]\times10^{12}$ and $V_{\text{disp}}\in[40,100]\,\text{meV}$, the noninteracting Fermi level cuts the lowest conduction band.
In this case, we find that the single-particle gap between the  noninteracting Fermi level and the remote bands is at least 5 times the characteristic scale of the Coulomb interaction $e^2 \sqrt{n_e}/(4\pi \epsilon_r \epsilon_0) $.
(See \appref{app:interaction-details}.)
Therefore, we can safely perform all calculations on the projected many-body model:
\eqa{
\label{eq:H_proj}
H_{\text{proj}} =
H_0 + H_1
 + \frac{1}{2 \A}\sum_{\bsl{q}}V(\bsl{q})
:\bar{\rho}_{\bsl{q}}\bar{\rho}_{-\bsl{q}}:   \ ,
}
where 
$
H_0 = \sum_{\bsl{k},\eta,s}\varepsilon_{\eta}(\bsl{k})
\gamma^\dagger_{\bsl{k},\eta,s}\gamma_{\bsl{k},\eta,s}
$
is the single-particle part, with $\varepsilon_{\eta}(\bsl{k})$ denoting the dispersion of the lowest conduction band in the valley $\eta$.
\(
\gamma_{\bsl{k},\eta,s}^\dagger
= \sum_{l,\sigma}U^{\eta }_{l\sigma}(\bsl{k})
c_{\bsl{k},l,\sigma,\eta,s}^\dagger
\) is the projected creation operator with $U^\eta(\bsl{k})$ the eigenvector for the lowest conduction band of the continuum Hamiltonian in valley $\eta$ for either spin.
\eq{
\bar{\rho}_{\bsl{q}} =
\sum_{\bsl{k},\eta,s}
M^{\eta}(\bsl{k},\bsl{q})
\gamma^\dagger_{\bsl{k}+\bsl{q},\eta,s}
\gamma_{\bsl{k},\eta,s},
} is the projected density, with the form factor given by
\eq{
M^{\eta}(\bsl{k},\bsl{q}) =
\sum_{l,\sigma}
U^{\eta *}_{l\sigma}(\bsl{k}+\bsl{q})
U^\eta_{l\sigma}(\bsl{k}).
}

The one-body term $H_1$ in \cref{eq:H_proj} depends on the choice of the interaction scheme \cite{Kwan2024MFCI-III,BAB2021TBGIII}.
There are two commonly adopted interaction schemes, the CN scheme and the average scheme.
Both schemes are studied in this work.
In the CN scheme, 
$H_1=0$, whereas in the average scheme, $H_1$ contains the Fock potential generated by the filled valence bands (in addition to a Hartree part that is proportional to the particle-number operator in this case due to the absence of a moir\'e potential).
Further discussions of the projection and the two schemes are given in \appref{app:interaction-details}.
The two schemes give similar phase diagrams, as discussed in the next section. We therefore show the CN-scheme results in the main text and present the average-scheme results in \appref{app:average-scheme-result}.

The projected many-body Hamiltonian retains all symmetries of the unprojected model at nonzero displacement field, including charge-spin $\mathrm{U}(2)$ symmetry in each valley, time-reversal symmetry, and the spinless threefold rotation $C_3$ about the out-of-plane axis.
In the following, we show that the HF solutions can spontaneously break these symmetries, particularly the spinless $C_3$ rotational symmetry in the nWC.

\section{HF phase diagram and nematic WC}
\label{sec:nematic-wigner-crystals}

\begin{figure*}
    \centering
    \IfFileExists{figures/phase_diagram_cn/hf_phase_diagram.pdf}{\includegraphics[width=0.8\textwidth]{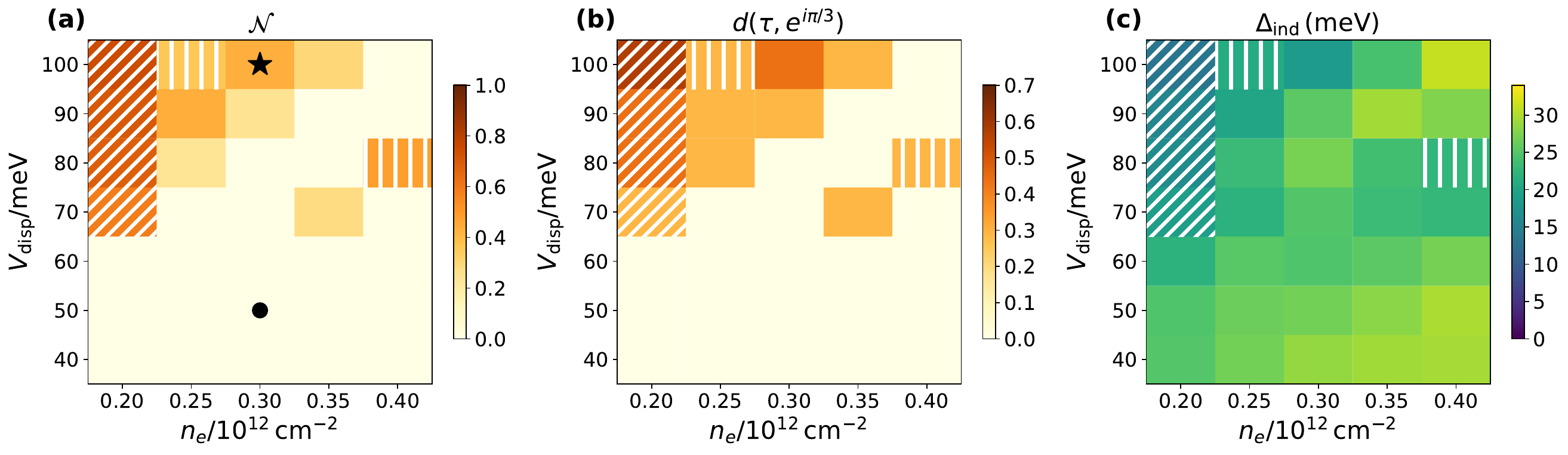}}{\fbox{\begin{minipage}[c][0.14\textheight][c]{0.8\textwidth}\centering phase diagram plots plot\end{minipage}}}
    \caption{
        HF phase diagram in the CN scheme for R$4$G.
        Vertical hatch indicates instability towards WCs with two electrons per unit cell.
        Diagonal hatch indicates instability towards metallic WCs, Fermi liquids, or stripe phases.
	    (a) $C_3$-symmetry violation in the density matrix, quantified by $\mathcal{N}$ as defined in \cref{eq:nematicity}. For $C_3$-invariant WCs, $\mathcal{N}=0$.
        The star is at $n_e=0.3\times10^{12}\,\text{cm}^{-2}$ and $V_\text{disp}=100\,\text{meV}$, and the circle is at $n_e=0.3\times10^{12}\,\text{cm}^{-2}$ and $V_\text{disp}=50\,\text{meV}$. 
	    (b) Hyperbolic lattice distance from the Wigner lattice of the HF ground state to the triangular lattice of the $C_3$-invariant WC, $d(\tau,e^{i\pi/3})$. For $\tau$ representing a $C_3$-invariant Wigner lattice, $d(\tau,e^{i\pi/3})=0$.
    	(c) HF indirect charge gap, which is greater than $10\,\text{meV}$ throughout the phase diagram.}
    \label{fig:cn_phase_diagram}
\end{figure*}

\begin{figure}[t]
\includegraphics[width=0.99\columnwidth]{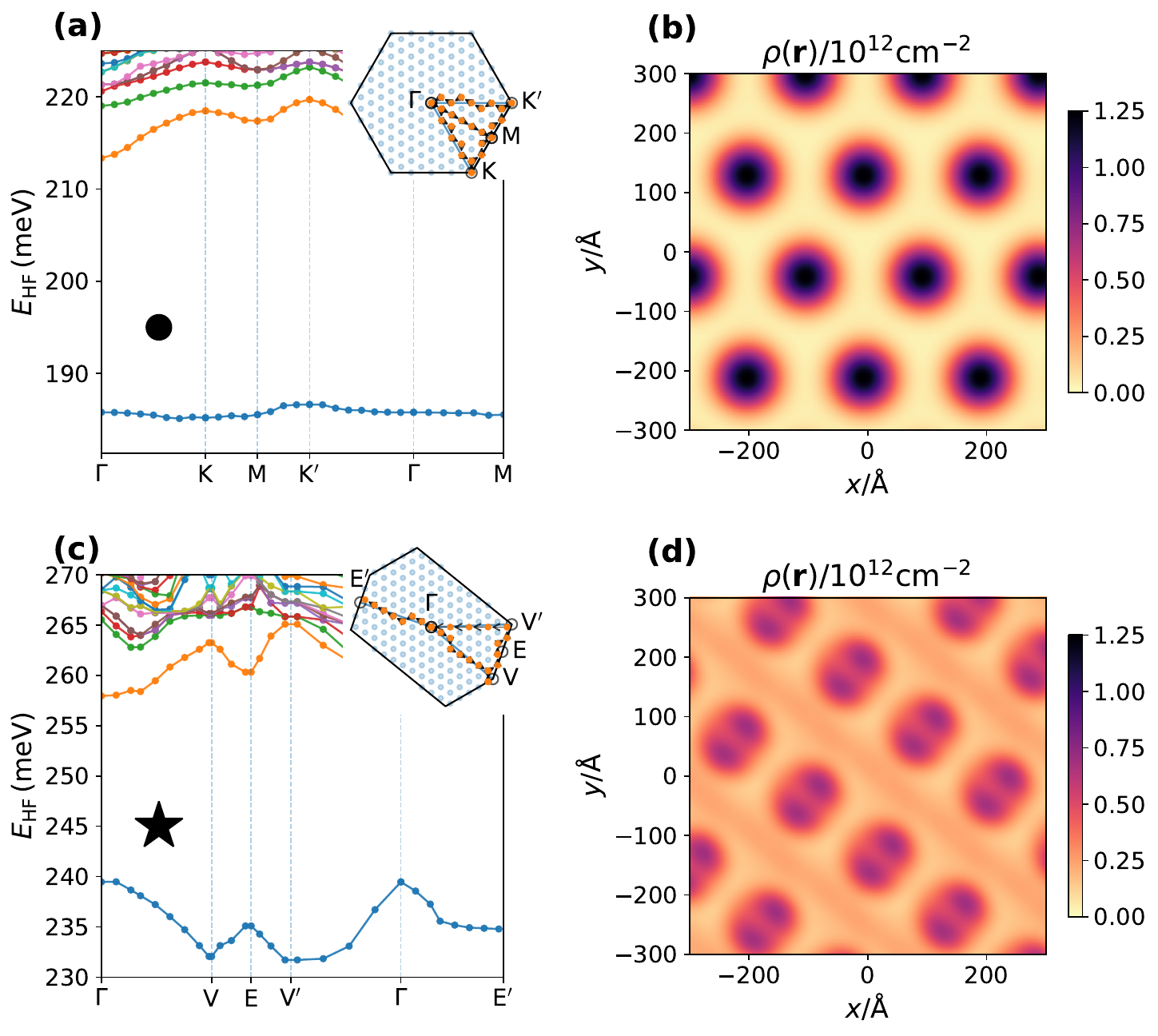}
\caption{
    (a),(c) Band structures and (b),(d) real-space charge densities for the representative $C_3$-invariant WC [the circle in \cref{fig:cn_phase_diagram}(a), displayed in (a) and (b) here], and nWC [the star in \cref{fig:cn_phase_diagram}(a), displayed in (c) and (d) here], respectively.
    The insets of (a) and (c) show the first Wigner Brillouin zone within the discretized continuum $k$ mesh. The rugged dispersion in (c) arises from imperfect commensuration between the discretized continuum $k$ mesh and the Wigner Bloch momentum points.
}
\label{fig:wc_representative_states}
\end{figure}

To perform the HF calculation, we first choose a momentum mesh in the continuum generated by the integer combinations of $\mathbf{B}_1=B(\sqrt{3},-1)/2$ and $\mathbf{B}_2=B(0, 1)$, where $B= 2 \sqrt{2}\pi / (3^{1/4}\sqrt{\A})$.
The mesh extends up to a cutoff $|\bsl{k}|\leq k_{\max}$ determined separately for each density and $V_\text{disp}$ from ground-state convergence, as detailed in \appref{app:cn:setup}.
Once the mesh is chosen, we perform HF calculations starting from tens of different random initial states. 
For each random initial state, we enforce one specific Wigner-lattice translational symmetry, which is characterized by two primitive Wigner reciprocal lattice vectors $\mathbf{b}_1$ and $\mathbf{b}_2$.
We make sure that $\mathbf{b}_1$ and $\mathbf{b}_2$ are always integer combinations of $\mathbf{B}_1$ and $\mathbf{B}_2$:
$(\mathbf{b}_1\ \mathbf{b}_2)=(\mathbf{B}_1\ \mathbf{B}_2) M$,
where $M$ is an integer matrix and $\det(M)=N_e$ with $N_e$ the number of conduction electrons.
Here $\det(M)=N_e$ means that there is on average one electron per Wigner unit cell.
At a fixed density and displacement field, varying $M$ yields converged HF states with different WC unit cells, and we identify the lowest-energy converged state as the ground state.
We emphasize that this direct comparison between different WC solutions is well-defined within our method because all solutions are constructed on the same underlying momentum mesh at a fixed density and displacement field.
This common mesh ensures that the resulting energy differences reflect the competition between distinct WC states, rather than artifacts arising from different momentum discretizations.

We benchmark the spin-valley polarization and the similarity between the CN and average schemes before performing large-size calculations.
To do so, we compute the HF phase diagrams for $N_e=36$ under the CN scheme (see \appref{app:cn-result}) and the average scheme (see \appref{app:average-scheme-result}). The two phase diagrams are similar.

Specifically, we find that for both schemes, (i) the HF ground state is always spin-valley polarized, (ii) the nWC states mostly appear at relatively high displacement fields and low densities in the phase diagram, as well as in a smaller region of intermediate fields and higher densities, and (iii) the Chern number is zero throughout the phase diagram.
Henceforth, we will focus on the CN scheme and examine the phase diagram at the larger system size of $N_e=144$ using spin-valley-polarized initial states.

The resultant phase diagram is shown in \Cref{fig:cn_phase_diagram}.
It shows that all the WCs considered are insulating and have an indirect gap $\Delta_{\text{ind}}>10\,\text{meV}$.
To distinguish a $C_3$-invariant WC from an nWC,
we introduce two complementary diagnostics of nematicity.
The first measures the asymmetry of a density matrix under spinless $C_3$ rotation,
\eq{
\label{eq:nematicity}
\mathcal{N}\equiv \underset{\bsl{r}}{\min} \big\| T_{\bsl{r}} C_3 P C_3^{\dagger} T_{\bsl{r}}^{\dagger}-P \big\|_{\max},
}
where $T_{\bsl{r}}$ accounts for real-space translations, and $\|P\|_{\max}\equiv\max_{ij}|P_{ij}|$ is the elementwise maximum norm.
The operator $T_{\bsl{r}}$ accounts for an arbitrary displacement of the WC lattice site from the spatial origin of the $C_3$ rotation.
No valley $\mathrm{U}(1)$ rotation is needed because our ground states are valley polarized.

Another diagnostic of nematicity is the hyperbolic lattice distance,
$d(\tau,e^{i\pi/3})$, defined as the minimum hyperbolic distance \cite{DHoker_2022_Modular} from the lattice of a $C_3$-invariant WC, \ie, $e^{i\pi/3}$, to
$\tau=b_2/b_1$,
where $b_{i}=(\mathbf{b}_{i})_1+\ii (\mathbf{b}_{i})_2$ is the complexified primitive reciprocal lattice vector and the minimum is taken over all primitive bases of the lattice. A $C_3$-invariant WC has zero hyperbolic lattice distance.
See \appref{app:hyperbolic-distance} for details.

Both diagnostics identify the same phase boundary in \cref{fig:cn_phase_diagram}(a) and (b).
To further visualize the two types of WCs, we plot the representative band structures and  real-space densities in \cref{fig:wc_representative_states}.
The real-space density in \cref{fig:wc_representative_states}(d) clearly illustrates the breaking of $C_3$ symmetry in the nWC, in contrast to the perfectly $C_3$-invariant density in \cref{fig:wc_representative_states}(b).
Both band structures are gapped.
Despite the ubiquitous indirect charge gap, some parts of the phase diagram in \cref{fig:cn_phase_diagram} are unstable, as revealed by TDHF below.

\section{TDHF and Stability of WCs}
\label{sec:tdhf-stability}

To assess the stability of the HF states, we use TDHF calculations to obtain approximated neutral excitations of a chosen HF state
\cite{Kwan2024MFCI-III,Rowe1970NuclearCollectiveMotion,RingSchuck1980NuclearManyBody,BlaizotRipka1986FiniteSystems}.
(See details in \appref{app:tdhf-details}.)
In particular, from linearized TDHF we construct the stability matrix $\mathcal{S}$
\cite{Thouless1960,Thouless1961,Nakada2016,Nakada2016add,Cui2013}, which determines the local stability of the HF solution with respect to particle-hole Thouless rotations at arbitrary Wigner Bloch momentum transfer $\bsl{q}$ \cite{Thouless1960}.
A locally stable HF solution has a positive-semidefinite stability matrix with zero eigenvalues that occur only at zero Wigner Bloch momentum transfer $\bsl{q}=0$ as Goldstone modes associated with spontaneously broken continuous symmetries (\eg, phonons arising from broken continuous translational symmetry).
If the stability matrix has nonpositive eigenvalues at a nonzero Wigner Bloch momentum transfer $\bsl{q}\ne0$, then there is an instability toward a WC with a different Wigner unit cell at a possibly different integer filling per cell, or even toward a state that is no longer a WC.
If the new Wigner unit cell does not allow integer filling for the same electron number, the resulting state would be a metallic WC.
If the new translation-breaking pattern is not a WC, the instability may instead lead to a Fermi liquid or a stripe phase.

\begin{figure}[t]
\centering
\begin{minipage}{\columnwidth}
\centering
\begin{tabular}{@{}cc@{}}
    \begin{tabular}{@{}l@{}}\textbf{(a)} \raisebox{-0.17em}{\Large{$\bullet$}}\\[-0.5ex]\IfFileExists{figures/tdhf_cn/tdhf_stability_c3.pdf}{\includegraphics[width=0.48\textwidth]{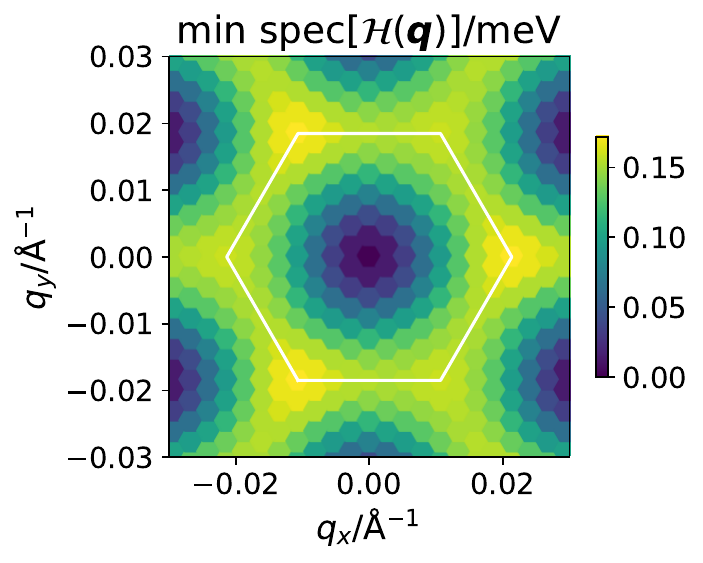}}{\fbox{\begin{minipage}[c][0.14\textheight][c]{0.5\textwidth}\centering $C_3$ TDHF stability spectra\end{minipage}}}\end{tabular} &
    \begin{tabular}{@{}l@{}}\textbf{(b)} $\bigstar$\\[-0.5ex]\IfFileExists{figures/tdhf_cn/tdhf_stability_nematic.pdf}{\includegraphics[width=0.49\textwidth]{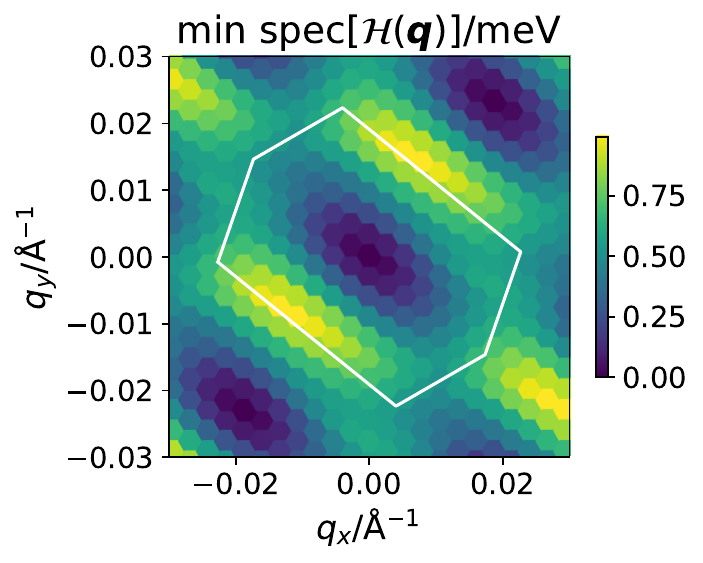}}{\fbox{\begin{minipage}[c][0.14\textheight][c]{0.46\textwidth}\centering Nematic TDHF stability spectra\end{minipage}}}\end{tabular}
\end{tabular}
\end{minipage} 
\caption{Lowest eigenvalue branch of the TDHF stability matrix for (a) a $C_3$-invariant WC [the circle in \cref{fig:cn_phase_diagram}(a)] and (b) an nWC [the star in \cref{fig:cn_phase_diagram}(a)]. The first Wigner Brillouin zones (BZs) are outlined in white. In both cases, the eigenvalues are positive at $\bsl{q}\neq 0$ and zero (within numerical error) at $\bsl{q}=0$, indicating local stability of the HF solution.
}
\label{fig:tdhf_stability}
\end{figure}

Based on our TDHF results, all $C_3$-invariant WC solutions and half of the nWC solutions are stable, as shown in \cref{fig:cn_phase_diagram}.
Representative eigenvalue maps of the stability matrix for the $C_3$-invariant WC and the nWC are shown in \cref{fig:tdhf_stability}(a) and (b), respectively, where the only zero eigenvalues occur at zero Wigner Bloch momentum transfer.
All unstable WC states are nematic (shown as hatched areas in \cref{fig:cn_phase_diagram}). 
Two of the unstable cases have clear instabilities at half of a Wigner reciprocal primitive vector, and thus tend to double their Wigner unit cell size to form a new WC with two electrons per unit cell~\cite{Moulopoulos_1992_Paired_Crystal,Moulopoulos_1993_Paired_Theory,Taut_1994_No_Paired,Zverevich_2026_Triplet_Wigner,MacDonald_2026_RMG_Guide}.
The others are unstable towards metallic WCs, Fermi liquids, stripe phases or other phases in general~\cite{Han_2026_Wigner_R4G,Nguyen_2026_CDW_SC,Zhou_2026_RMG_WC,Dong_2026_mWC_RMG,Zaletel_2026_mWC_RMG}.

\section{Kinetic Energy Gain of the nWC}
\label{sec:nematic-mechanism}

\begin{figure*}[t]
    \centering
    \begin{tabular}{ccc}
    \begin{tabular}{@{}l@{}}\textbf{(a)}\\[-0.5ex]\includegraphics[width=0.257\textwidth]{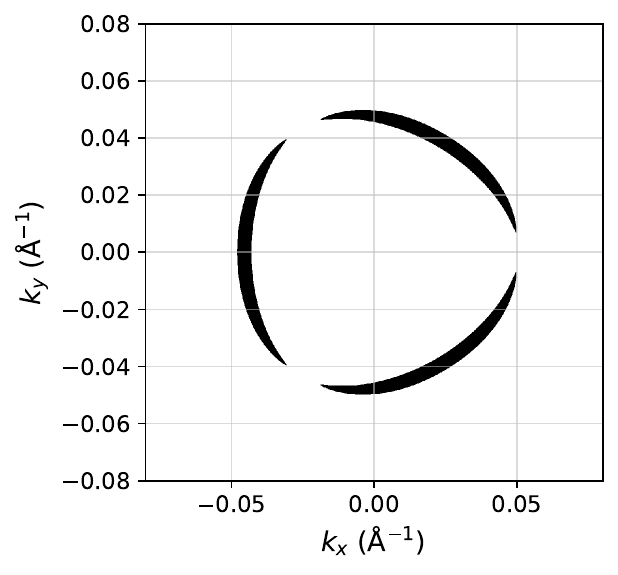}\end{tabular} &
    \begin{tabular}{@{}l@{}}\textbf{(b)}\\[-0.5ex]\includegraphics[width=0.3\textwidth]{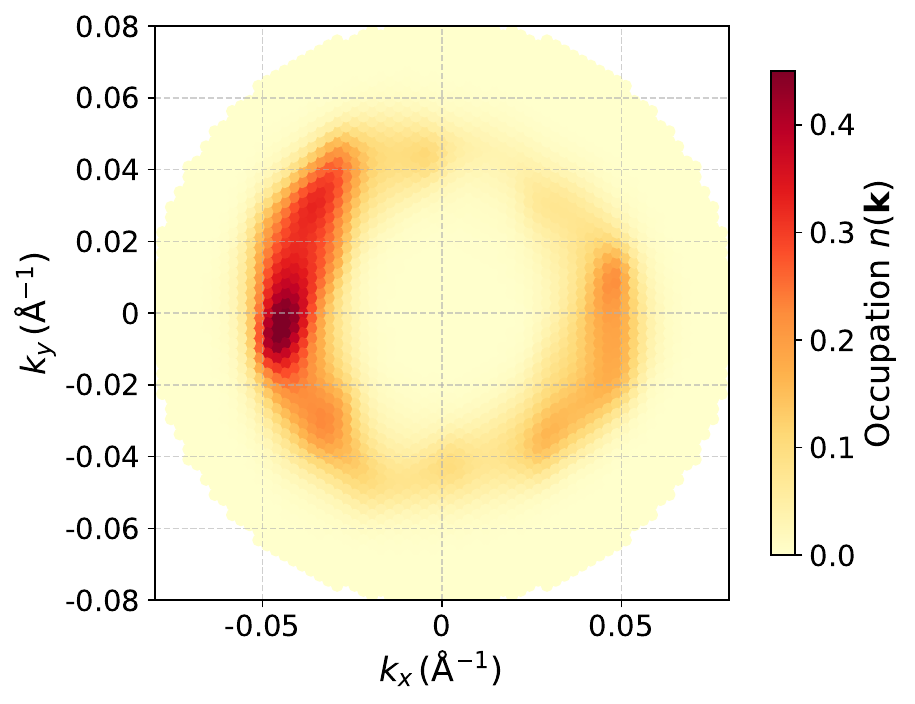}\end{tabular} &
    \begin{tabular}{@{}l@{}}\textbf{(c)}\\[-0.5ex]\includegraphics[width=0.305\textwidth]{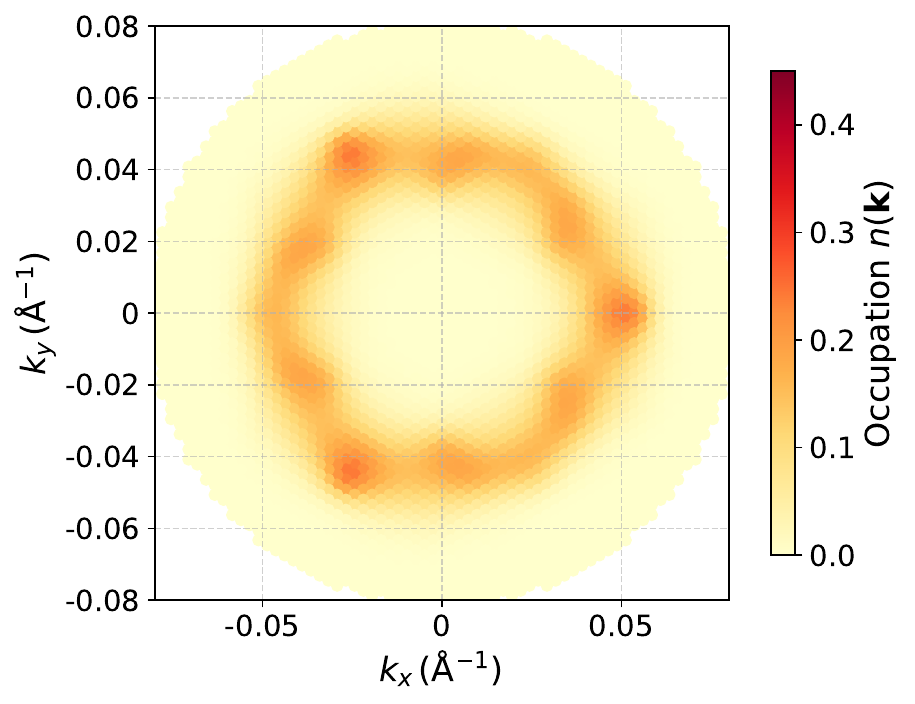}\end{tabular}
    \end{tabular}
    \caption{Momentum-space origin of the nWC at $n_e=0.25\times10^{12}\,\text{cm}^{-2}$ and $V_\text{disp}=90\,\text{meV}$. (a) Noninteracting Fermi sea of a spin-valley-polarized state, showing the three-lobed low-density Fermi surface (in black) produced by trigonal warping. (b) Occupation $n(\bsl{k})$ of the nWC ground state.
    (c) Occupation $n(\bsl{k})$ of the lowest-energy $C_3$-invariant WC  solution at the same parameters. This higher-energy solution preserves the threefold occupation pattern and spreads weight over the three lobes.
    }
    \label{fig:nematic_mechanism}
\end{figure*}

We now discuss how the nWC outcompetes the $C_3$-invariant WC in the larger region of large displacement fields and small densities.
Compared with the noninteracting Fermi sea, any crystalline state necessarily incurs a kinetic-energy cost, which means they must be stabilized by lower correlation energies.

Nevertheless, comparing an nWC with an $C_3$-invariant WC derived from the same noninteracting Fermi sea does not immediately reveal whether nematicity is favored by a lower kinetic or correlation energy.
To clarify this, we compare the kinetic and correlation energies of the HF ground state in the CN scheme with those of the lowest-energy $C_3$-invariant solution in the nWC phase. 
For example, at $n_e=0.25\times10^{12}\,\text{cm}^{-2}$ and $V_\text{disp}=90\,$meV, the nWC has a kinetic energy per electron that is $0.328\,$meV lower than that of the $C_3$-invariant solution, while its correlation energy per electron is $0.221\,$meV higher.
Thus, the nWC is favored by its kinetic-energy gain.
In fact, this is true for almost all of the nWC phase, with the only exception being the stable outlier nWC in the phase diagram: $n_e=0.4\times10^{12}\,\text{cm}^{-2}$ and $V_\text{disp}=80\,$meV. (See \appref{app:cn-result}.)

The kinetic energy gain can also be seen in the momentum occupation plot.
Figure~\ref{fig:nematic_mechanism}(a) shows the noninteracting Fermi sea at $n_e=0.25\times10^{12}\,\text{cm}^{-2}$ and $V_\text{disp}=90\,\text{meV}$, where the HF ground state is a stable nWC.
The single-particle Fermi sea is strongly distorted by trigonal warping and is therefore far from circular.
Instead, it consists of three disconnected arc-like lobes related to one another by $C_3$ rotation.
As shown in \cref{fig:nematic_mechanism}(b) and (c), the momentum distribution $n(\bsl{k})$ of the nWC is predominantly concentrated around one of the three lobes, while the electronic occupation of the $C_3$-invariant WC is spread over all three lobes.
In particular, the occupation of the nWC state inside the noninteracting Fermi sea is 1.78\% higher than that of the $C_3$-invariant solution.
Therefore, in the CN scheme the nWC involves fewer high-kinetic-energy states than the $C_3$-invariant WC, leading to spontaneous nematicity.
Kinetically favored nWCs (compared to $C_3$-invariant WCs) also appear in the average interaction scheme. Thus this kinetic energy gain is not an artifact of a specific interaction scheme (see \appref{app:average-scheme-result}).
A more detailed physical intuition and analytic understanding of the mechanism underlying the nWC formation are left for future work.

\section{Conclusion and Discussion}
\label{sec:discussion}

In conclusion, our HF and TDHF results show that the WC phase in R4G hosts a stable spin-valley-polarized nWC that spontaneously breaks the spinless threefold rotational symmetry.
At low $V_{\text{disp}}$ or high $n_e$, almost all WCs are in a triangular $C_3$-invariant state, while at high $V_{\text{disp}}$ and low $n_e$, the Wigner lattice lowers its energy by distorting in a way that favors momentum occupation in one of the three symmetry-related directions.

Our proposed nWC can be tested by angle-resolved transport measurements in a ``sunflower'' device geometry~\cite{Zhang_2024_ARTM_Sunflower,Zhang_2022_ElectronicAnisotropy_TTG,Morissette_2025_Nematicity_R6G,Qin_2025_Anisotropy_RMG}, in which many contacts surround the sample and the current direction can be rotated within the plane.
In such a setup, the nWC should show an anisotropic critical voltage or depinning threshold: driving along the easy and hard axes of the pinned crystal should require different voltages to destabilize the insulating state.
Although we focus on R4G, we would expect nWCs to occur in other rhombohedral multilayers, as long as the three-lobed Fermi sea exists.
We have not considered the effect of disorder in this work, but it is an important future direction to include disorder in the calculations to distinguish the Anderson solid from the Wigner solid in the phase diagram \cite{Huang_2026_disorder_2D,Babbar_2026_Wigner_Anderson,Babbar_2026_2D_Transport}.

\begin{acknowledgments}
We thank B.~Andrei Bernevig, Sankar Das Sarma, Jonah Herzog-Arbeitman, Long Ju, Yves Kwan, and J.I.A.\ Li for helpful discussions.
This work is supported by startup funds at the University of Florida. The authors acknowledge UFIT Research Computing for providing computational resources and support. Z.L. is supported by startup funds from Florida State University. 
\end{acknowledgments}

\bibliography{bibfile_references,bib2,bib3}

\appendix
\crefname{section}{App.}{Apps.}
\Crefname{section}{Appendix}{Appendices}

\onecolumngrid

\newpage
\tableofcontents

\section{Single-particle Models}\label{app:DFT}

\begin{figure}[htbp]
\centering
\includegraphics[width=0.8\linewidth]{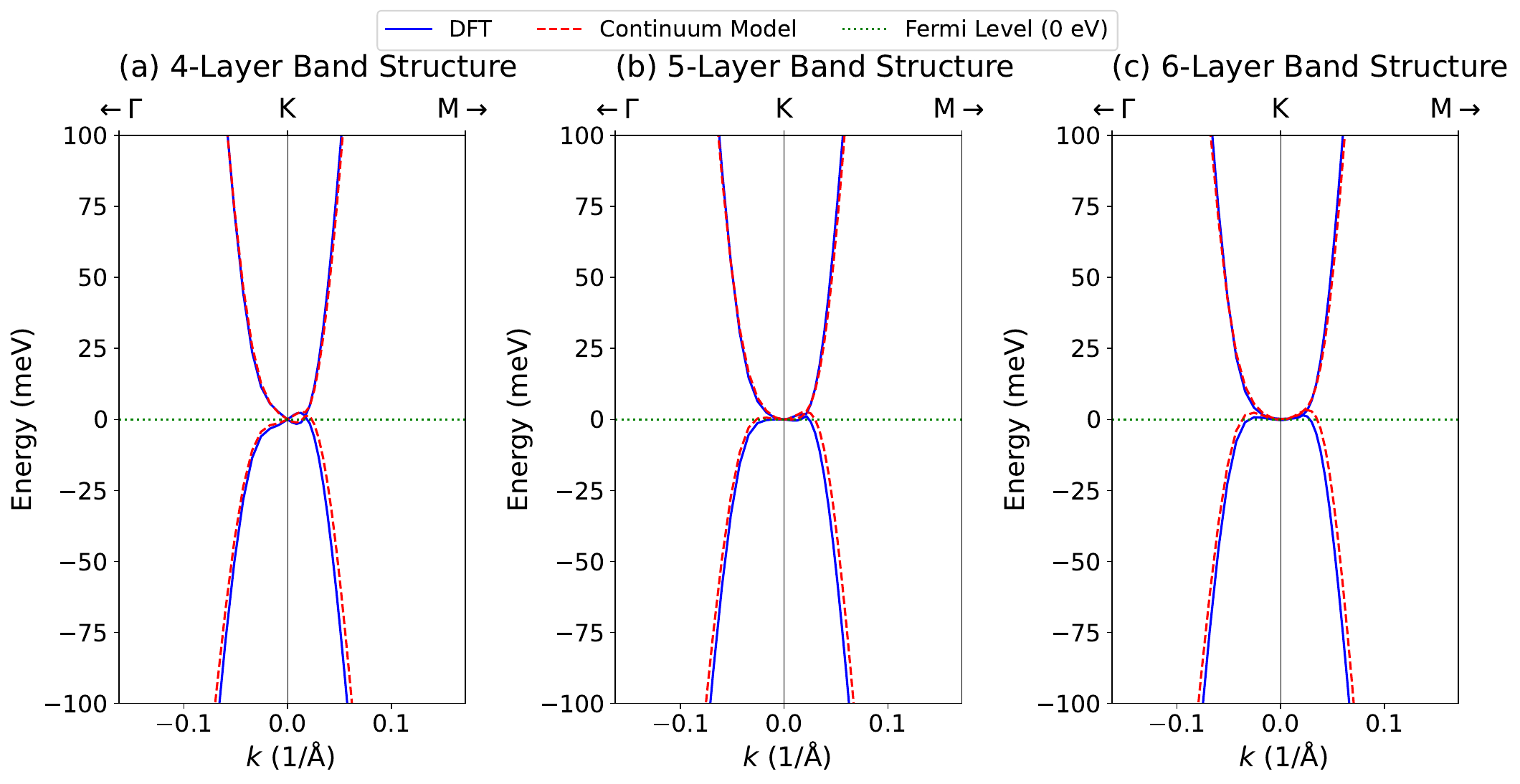}
\caption{Comparison of the low-energy electronic band structures calculated using density functional theory (solid blue lines) and the continuum model (dashed red lines). The panels show the bands for (a) four-layer, (b) five-layer, and (c) six-layer systems near the $K$ valley.
The $k$-path is part of the $\Gamma$--$K$--$M$ trace that extends from $K$ by 9.55\% of the distance towards $\Gamma$, and by 20.1\% of the distance towards $M$.
The Fermi level is aligned at zero energy (green dotted line) for both models.
}
\label{fig:band_comparison}
\end{figure}
 
In this work, the single-particle part of the model is adopted from \refcite{herzog2024moire} with $v_F = 542.1\,$meV$\cdot$nm, $v_3 = v_4 = 34.0\,$meV$\cdot$nm, $t_1 = 355.16\,$meV, $t_2 = -7\,$meV, and $V_{\text{ISP}} = 16.65\,$meV.
\cref{fig:band_comparison} presents a comparison of the electronic band structures for rhombohedral 4-, 5-, and 6-layer graphene, calculated using the low-energy continuum model (dashed red lines) and density functional theory (DFT, solid blue lines) at $V_\text{disp}=0$ in the $K$ valley.
As shown in the subpanels, within the low-energy window ($\pm 100$~meV) surrounding the Fermi level, there is good quantitative agreement between the two methods. 
In particular, the continuum model accurately reproduces the asymmetric band dispersion along the $k$-path shown,
which extends from $K$ by 9.55\% of the distance toward $\Gamma$ and by 20.1\% of the distance toward $M$, confirming that trigonal-warping effects and interlayer hopping parameters are properly captured.
This agreement confirms that the parameterized low-energy continuum model accurately describes the single-particle electronic properties of these multilayer systems.

\section{Interaction Hamiltonian Details} \label{app:interaction-details}

In the absence of interactions, the electron density in the WC phase is sufficiently low that only the lowest conduction band is occupied.
Furthermore, the characteristic interaction scale in a WC is smaller than the interband excitation-energy scale throughout the parameter regime considered.
We estimate the interaction energy using $V\sim \frac{e^2 \sqrt{n_e}}{4\pi\epsilon_0\epsilon_r}$ and compare it to the interband particle or hole excitation energy, which is the energy difference between the Fermi surface and the bottom of the next conduction band or between the Fermi surface and the top of the highest valence band.
The interaction energy is less than $1/5$ of these interband excitation energies throughout the phase diagram.
This justifies projecting the Hamiltonian onto the lowest conduction band.
Different schemes for projecting the interaction terms in HF can produce different physics.

In this section, we review two different interaction schemes used in the projected HF calculations, following \refcite{Kwan2024MFCI-III}.
In a projected HF calculation, one selects an active subspace $\mathscr{H}_{\text{act.}}$ of the full Hilbert space $\mathscr{H}$ and optimizes the components of the Slater determinant within $\mathscr{H}_{\text{act.}}$.
Although the occupations in the rest of $\mathscr{H}$ are frozen, the frozen bands can still contribute to the projected HF Hamiltonian in $\mathscr{H}_{\text{act.}}$ through interaction terms.
Different interaction schemes dictate the form of this contribution.

In our model, the active subspace is the lowest conduction band of R4G. In the charge-neutrality (CN) scheme, the background charge is assumed to screen the Coulomb field of the valence bands.
In contrast, the average scheme assumes an infinite-temperature reference and accounts for the Coulomb potential due to the valence-band electrons through their form factors.

\subsection{General formalism of the projection and interaction schemes}

The main assumption of the general projection is that the energetically remote bands, \eg, those that are far away from charge neutrality, can be considered frozen to be either completely filled or empty.
Thus, we project the problem into an active subspace described by a set of single-particle states $\mathscr{H}_{\text{act.}}$ within the full Hilbert space $\mathscr{H}$:
\eq{
\mathscr H
=
\mathscr H_{\text{rem.val.}}
\oplus
\mathscr H_{\text{act.}}
\oplus
\mathscr H_{\text{rem.cond.}},
\label{eq:app_hilbert_decomp}
}
where ``rem.val.'' and ``rem.cond.'' denote the completely filled remote valence bands and the empty remote conduction bands, respectively.
For the study in the main text, we divide the Hilbert space using the spectrum of the single-particle Hamiltonian, $H_0$, and choose $\mathscr{H}_{\text{act.}}$ to be spanned by the first conduction band in R4G.

Let $L_{\text{act.}}$ be the space of linear operators on $\mathscr{H}_{\text{act.}}$. To construct an active Hamiltonian ${H}_{\text{act.}}\in L_{\text{act.}}$ that captures the physics, we need to take into account the effect of the fully occupied remote valence bands. 
This is done by normal ordering the bare interaction $H_{\text{int}}$ with respect to the completely filled $\mathscr H_{\text{rem.val.}}$ and completely empty $\mathscr{H}_{\text{act.}}\oplus \mathscr H_{\text{rem.cond.}}$:
\eq{
{H}_{\text{int}} = :{H}_{\text{int}}: + {H}_{\text{rem.val.}}^{\text{eff}}\ ,
}
where ``bare'' means that the interaction is originally normal-ordered with respect to the vacuum of all electrons, $:...:$ denotes normal ordering, and ${H}_{\text{rem.val.}}^{\text{eff}}$ captures the additional one-body term generated by normal ordering.
In our work, ${:{H}_{\text{int}}:}$ simplifies to the familiar form of normal ordering of the screened Coulomb interaction with respect to the conduction vacuum.
Since ${H}_{\text{rem.val.}}^{\text{eff}}$ comes from the completely filled valence bands, it takes the following form:
\eq{
    {H}_{\text{rem.val.}}^{\text{eff}} \equiv {H}_{\text{HF,int}}[P_{\text{rem.val.}}] \Big|_{\text{act.}} ,
}
where $P_{\text{rem.val.}} = I_{\text{rem.val.}}$ is the identity in the remote valence subspace.
${H}_{\text{HF,int}}[P]$ denotes the Hartree and Fock interaction terms generated by a density matrix $P$.
The full projected Hamiltonian is then built with an additional reference density matrix $P^{\text{ref}}$ in the full Hilbert space:
\eqa{
\label{eq:app:H_act}
    {H}_{\text{act.}} & = {H}_0\Big|_{\text{act.}} + {:{H}_{\text{int}}:}\Big|_{\text{act.}}
     + \left({H}_{\text{rem.val.}}^{\text{eff}}
     - {H}_{\text{HF,int}}[P^{\text{ref}}]\right)\Big|_{\text{act.}} \\
     & = {H}_0\Big|_{\text{act.}} + {:{H}_{\text{int}}:}\Big|_{\text{act.}}
     + {H}_{\text{HF,int}}[P_{\text{rem.val.}}-P^{\text{ref}}]\Big|_{\text{act.}} \\
     & = {H}_0\Big|_{\text{act.}} + {:{H}_{\text{int}}:}\Big|_{\text{act.}}
     + {H}_1\ ,
}
where ``$|_{\text{act.}}$'' represents the truncation to terms within $L_{\text{act.}}$, and 
\eq{
{H}_1 ={H}_{\text{HF,int}}[P_{\text{rem.val.}}-P^{\text{ref}}]\Big|_{\text{act.}}\ .
}
Different interaction schemes are defined by different choices of $P^{\text{ref}}$, or equivalently different ${H}_1$, to capture background charge and screening effects.

\subsection{The CN and average schemes}

In the CN scheme,
$P^{\text{ref.CN}} = I_{\text{val.}}$ is the identity in the valence bands, which cancels ${H}_{\text{rem.val.}}^{\text{eff}}$ in \cref{eq:app:H_act}, rendering $H^{\text{CN}}_1=0$, and leaves only the normal-ordered interaction terms in $L_{\text{act.}}$, as in \cref{eq:H_proj}.

In the average scheme, an infinite-temperature reference density matrix is assumed for the background charge. The Coulomb interaction is written in terms of the density fluctuation above a uniform background, as $\delta{\rho}_{\bsl{q}}\delta{\rho}_{-\bsl{q}}$, where
\eq{
    \delta{\rho}_{\bsl{q}} = \sum_{\bsl{k}, l,\sigma,\eta,s}
    \left(
    {c}^\dagger_{\bsl{k+q}, l,\sigma,\eta,s} {c}_{\bsl{k}, l,\sigma,\eta,s}-
    \frac{1}{2}\delta_{\bsl{q},\bsl{0}}
    \right), 
}
where ${c}_{\bsl{k}, l,\sigma,\eta,s}$ is the annihilation operator for a plane wave state with momentum $\bsl{k}$, on sublattice $\sigma$, in layer $l$, valley $\eta$, and with spin $s$.
Transforming to the complete band basis of $H_0$,
\eq{
    {c}_{\bsl{k}, l,\sigma,\eta, s} =\sum_{n} U^\eta_{n,l,\sigma}(\bsl{k}) \gamma_{n, \bsl{k},\eta,s},
}
where, compared with the main text, we have restored the band index $n$, which runs over all bands, and $U^{\eta}_{n,l,\sigma}(\bsl{k})$ denotes the eigenvector of the $n$th band of $H_0$ in valley $\eta$ with spin $s$.
The form factors in the band basis are
\eq{
    M^\eta_{m, n}(\bsl{k},\bsl{q}) = \sum_{l,\sigma}
    U^{\eta*}_{m,l,\sigma}(\bsl{k}+\bsl{q})
    U^\eta_{n,l,\sigma}(\bsl{k}).
}
By construction, $M^\eta_{m, n}(\bsl{p},\bsl{q}) = M^{\eta*}_{n, m}(\bsl{p}+\bsl{q},-\bsl{q})$, and the normalization is $M^\eta_{m,n}(\bsl{k},\bsl{0})=\delta_{mn}$.
Then, the density fluctuation in the band basis becomes
\eq{
  \delta {\rho}_{\bsl{q}} =
  \sum_{\bsl{p},\eta, m, n, s} M^\eta_{m,
  n} (\bsl{p}, \bsl{q}) \left(\gamma^\dagger_{m,\bsl{p}+\bsl{q},\eta,s} \gamma_{n,\bsl{p},\eta,s} - \frac{1}{2}
  \delta_{\bsl{q}, \bsl{0}} \delta_{m n} \right).
}
Let $\mathcal{A} = (2\pi)^2/\mathrm{d}^2k$ be the sample area.
The interaction Hamiltonian is
\begin{eqnarray}
  {H}^{\text{avg}}_{\text{int}} & = & 
  \frac{1}{2\mathcal{A}}\sum_{\bsl{q}}V(\bsl{q})\delta{\rho}_{\bsl{q}}\delta{\rho}_{-\bsl{q}}
  \nonumber \\
  &=& {H}^{\text{avg}\prime}_{\text{int}} + \frac{1}{2\mathcal{A}}\sum_{\bsl{q}} V (\bsl{q}) \sum_{m\bsl{k}\eta{s}}
  \gamma_{m\bsl{k}\eta{s}}^{\dag} \gamma_{m\bsl{k}\eta{s}}
  -  \frac{V (\bsl{0})}{2\mathcal{A}} \sum_{m\bsl{k}\eta{s}} \gamma_{m\bsl{k}\eta{s}}^{\dag}
  \gamma_{m\bsl{k}\eta{s}} \Big( \sum_{n\bsl{p}\eta'{s'}} 1 \Big) + \frac{V
  (\bsl{0})}{8\mathcal{A}} \Big( \sum_{n\bsl{p}\eta{s}} 1 \Big)^2,
  \end{eqnarray}
where
\begin{eqnarray}
{H}^{\text{avg}\prime}_{\text{int}} & = & 
\frac{1}{2 \mathcal{A}} \sum_{
\begin{subarray}{c}
    \bsl{k}_1 \bsl{k}_2 \bsl{q}\\
    \eta_1 \eta_2 s_1 s_2
  \end{subarray}} 
  \Bigg( \sum_{m_1 m_2 m_3 m_4 \in \text{cond.}} 
  + 2 \sum_{
  \begin{subarray}{c}
    m_1 m_3 \in \text{cond.}\\
    m_2 m_4 \in \text{val.}
  \end{subarray}} 
  + 2 \sum_{
  \begin{subarray}{c}
    m_1 m_4 \in \text{cond.}\\
    m_2 m_3 \in \text{val.}
  \end{subarray}}
  \Bigg) \gamma_{m_1,\bsl{k}_1+\bsl{q},\eta_1,s_1}^{\dag} \gamma_{m_2, \bsl{k}_2 -\bsl{q},\eta_2,s_2}^{\dag}
  \gamma_{m_3 \bsl{k}_2 \eta_2 s_2} \gamma_{m_4 \bsl{k}_1 \eta_1 s_1}\nonumber\\
  &  & \times V (\bsl{q}) M^{\eta_1}_{m_1, m_4} (\bsl{k}_1,\bsl{q}) M^{\eta_2}_{m_2, m_3} (\bsl{k}_2, - \bsl{q}).
\end{eqnarray}
Terms other than $H^{\text{avg}\prime}_{\text{int}}$ are proportional to the number operator.
The terms mixing conduction and valence bands need further normal ordering.
\begin{eqnarray}
{H}^{\text{avg}\prime}_\text{int,ph-exchange}& \equiv &\frac{1}{ \mathcal{A}}\!
\sum_{
\begin{subarray}{c}
    \bsl{k}_1\bsl{k}_2\bsl{q}\\
    \eta_1\eta_2 s_1 s_2
  \end{subarray}
}
\sum_{
\begin{subarray}{c}
    m_1 m_3 \in \text{cond.}\\
    m_2 m_4 \in \text{val.}
  \end{subarray}
}
\gamma_{m_1, \bsl{k}_1
+\bsl{q},\eta_1,s_1}^{\dag} \gamma_{m_2, \bsl{k}_2 -\bsl{q},\eta_2,s_2}^{\dag}
\gamma_{m_3 \bsl{k}_2 \eta_2 s_2} \gamma_{m_4 \bsl{k}_1 \eta_1 s_1} V(\bsl{q}) M^{\eta_1}_{m_1, m_4} (\bsl{k}_1,\bsl{q}) M^{\eta_2}_{m_2, m_3} (\bsl{k}_2, - \bsl{q})\nonumber\\
& = &
:{H}^{\text{avg}\prime}_\text{int,ph-exchange}: \nonumber\\
&& - \frac{1}{\mathcal{A}} \sum_{\bsl{k} \eta s}  \sum_{m_1, m_3 \in \text{cond.}} \gamma_{m_1 \bsl{k} \eta s}^{\dag} \gamma_{m_3 \bsl{k} \eta s} \sum_{\bsl{q}} V (\bsl{q})
\sum_{m_2 \in \text{val.}} M^{\eta*}_{m_2, m_1} (\bsl{k}, - \bsl{q}) M^\eta_{m_2, m_3} (\bsl{k}, -
\bsl{q}) ,\\
{H}^{\text{avg}\prime}_\text{int,ph-direct}
&\equiv&
\frac{1}{ \mathcal{A}}\!
\sum_{
\begin{subarray}{c}
    \bsl{k}_1\bsl{k}_2\bsl{q}\\
    \eta_1\eta_2 s_1 s_2
  \end{subarray}
}
\sum_{
\begin{subarray}{c}
    m_1 m_4 \in \text{cond.}\\
    m_2 m_3 \in \text{val.}
  \end{subarray}
}
\gamma_{m_1, \bsl{k}_1
+\bsl{q},\eta_1,s_1}^{\dag} \gamma_{m_2, \bsl{k}_2 -\bsl{q},\eta_2,s_2}^{\dag}
\gamma_{m_3 \bsl{k}_2 \eta_2 s_2} \gamma_{m_4 \bsl{k}_1 \eta_1 s_1} V(\bsl{q}) M^{\eta_1}_{m_1, m_4} (\bsl{k}_1,\bsl{q}) M^{\eta_2}_{m_2, m_3} (\bsl{k}_2, - \bsl{q})\nonumber\\
& = & 
:{H}^{\text{avg}\prime}_\text{int,ph-direct}: + \frac{V(\bsl{0})}{ \mathcal{A}}\sum_{\bsl{k}\eta s} \Big( \sum_{m \in \text{cond.}} \gamma_{m
\bsl{k}\eta s}^{\dag} \gamma_{m \bsl{k}\eta s} \Big)  \Big( \sum_{\bsl{p}\eta' s'} \sum_{n \in \text{val.}} 1
\Big).\label{eq:app:ref-hartree}
\end{eqnarray}
Recall that normal ordering places creation (annihilation) operators for valence-band (conduction-band) states on the right.
Since the valence band is filled and frozen, both normal-ordered terms in ${H}^{\text{avg}\prime}_\text{int,ph-exchange}$ and ${H}^{\text{avg}\prime}_\text{int,ph-direct}$ vanish.
Therefore,
\begin{eqnarray}
{H}^{\text{avg}}_\text{int} &=& :{H}^{\text{avg}}_\text{int}: + {H}^{\text{avg}}_1 + \text{(constant and number operator)}, \\
{H}^{\text{avg}}_1 &=& - \frac{1}{\mathcal{A}} \sum_{\bsl{k}\eta s} \sum_{m, n \in \text{cond.}} \gamma_{m \bsl{k}\eta s}^{\dag} \gamma_{n \bsl{k}\eta s} \sum_{\bsl{q}} V (-\bsl{q})
\sum_{l \in \text{val.}} M^{\eta*}_{l m} (\bsl{k}, \bsl{q}) M^\eta_{l n} (\bsl{k}, \bsl{q}) .
\label{eq:app:H_int_avg}
\end{eqnarray}
Assuming $V(\bsl{q})=V(-\bsl{q})$, one can see that ${H}^{\text{avg}}_1\in L_\text{act.}$ takes the form of the Fock potential generated by a completely filled valence band, \ie, by the identity matrix in the valence-band subspace.
The second term in \cref{eq:app:ref-hartree} is the Hartree term generated by the identity matrix in the valence-band subspace. Therefore, in the average scheme, $P^{\text{ref}}=0$.

\section{Diagnostics of Nematicity}

In the paper, we introduced two diagnostics of nematicity: the hyperbolic lattice distance $d(\tau,e^{i\pi/3})$ and the $C_3$ violation   of the density matrix, denoted by $\mathcal{N}$. They are further detailed below.

\subsection{Review of the hyperbolic distance \label{app:hyperbolic-distance}}

A 2D primitive lattice is spanned by a pair of linearly independent vectors 
$\mathbf{a}_1, \mathbf{a}_2 \in \mathbb{R}^2$. By representing these vectors as 
complex numbers $z_1=a_{1,x}+ia_{1,y}$ and $z_2 =a_{2,x}+ia_{2,y}$, the shape of the lattice, up to an overall scale and rotation, is completely 
determined by its modular parameter $\tau$, defined in the upper half-plane $\mathbb{H}=\{c\in\mathbb{C}:\operatorname{Im}(c)>0\}$ as their ratio
\eq{
\tau=\left\{\begin{array}{ll}
  z_2/z_1, & \operatorname{Im}(z_2/z_1)>0,\\
  z_1/z_2, & \text{otherwise.}
\end{array}\right.
}
$\operatorname{Im}(\tau)>0$ ensures that the basis is positively oriented.
To find the distance between two different lattices characterized by parameters 
$\tau_1$ and $\tau_2$, we use the natural Poincar\'e hyperbolic metric on $\mathbb{H}$. 
The hyperbolic distance $d_{\mathbb{H}}$ between them is given explicitly by \cite{DHoker_2022_Modular}
\eq{
d_{\mathbb{H}}(\tau_1, \tau_2) = \operatorname{arccosh}\left( 1 + \frac{|\tau_1 - \tau_2|^2}{2 \operatorname{Im}(\tau_1)  \operatorname{Im}(\tau_2)} \right)
\label{eq:hyperbolic_distance}
}
Primitive bases of the same lattice are related by unimodular transformations in $\mathrm{SL}(2,\mathbb{Z})$, which are generated by $(\mathbf{a}_1, \mathbf{a}_2)\to(\mathbf{a}_1+\mathbf{a}_2, \mathbf{a}_2)$ and $(\mathbf{a}_1, \mathbf{a}_2)\to(-\mathbf{a}_2, \mathbf{a}_1)$.
An element of $\mathrm{SL}(2,\mathbb{Z})$, say $\gamma$, acts by the transformation: $\gamma\cdot\tau=(a\tau+b)/(c\tau+d)$ where $ad-bc=1$, and $a,b,c,d\in\mathbb{Z   }$.
Consequently, physically equivalent lattices correspond to the same modular orbit \cite{DHoker_2022_Modular}.
The true shape distance $d(\tau_1, \tau_2)$ between two lattices 
is then defined by the minimum distance over all physically equivalent representations:
\eq{
d(\tau_1, \tau_2) \equiv \min_{\gamma \in \mathrm{SL}(2, \mathbb{Z})} d_{\mathbb{H}}(\tau_1, \gamma \cdot \tau_2)
\label{eq:lattice_distance}
}
In particular, $d(\tau_{C_3},e^{i\pi/3})=0$ for any modular parameter $\tau_{C_3}$ that represents a triangular lattice.

\subsection{$C_3$ violation in the density matrix}
\label{app:nematicity}

To quantify the spontaneous breaking of the threefold rotational
symmetry, we introduce a gauge-independent measure based directly on the
HF one-body density matrix. Let $C_3$ be the spinless threefold rotation about the out-of-plane axis acting on
both crystal momentum and internal degrees of freedom. Its matrix elements in the layer-orbital subspace, with layer index \(l=1,\ldots,N_\ell\) and sublattice
\(\sigma=A,B\), are given by
\eq{
(C_3^\text{internal})^\eta_{l\sigma,l'\sigma'} = \delta_{ll'}\delta_{\sigma\sigma'} \exp\!\left\{ i\eta\pi \left[ 1+ \frac{2}{3} \left(l-1-\lfloor N_\ell/2 \rfloor\right) \right] \right\},
}
where \(\eta=+1\) for the \(K\) valley, while \(\eta=-1\) for the \(-K\) valley.

For the self-consistent HF density matrix $P$, we define $C_3$ violation as
\begin{equation}
\mathcal N \equiv \min_{\bsl{\theta}} \left\| g_{\bsl{\theta}} C_3 P C_3^{-1} g_{\bsl{\theta}}^{-1} - P \right\|_{\max},
\label{eq:C3_violation}
\end{equation}
where $g_{\bsl{\theta}}$ is a general gauge transformation parameterized by $\bsl{\theta}$, and $\|\cdot\|_{\max}$ denotes the maximum absolute value among all
entries of the matrix. Specifically, gauge transformations in general include two-dimensional real-space translations $T_{\bsl{r}}$ and valley-spin $\mathrm{U}(2)\times \mathrm{U}(2)$ rotations.
Since the ground states in our work are all spin- and valley-polarized, the valley-spin $\mathrm{U}(2)\times \mathrm{U}(2)$ rotations can be omitted because they have no effect.

Unlike the hyperbolic distance introduced above, $\mathcal N$ probes the full $C_3$ symmetry of the HF state. It is therefore sensitive to rotational
symmetry breaking originating from either the Wigner lattice geometry or
the internal electronic structure encoded in the density matrix even if the Wigner lattice is $C_3$-invariant.

\section{Charge neutrality scheme: methods and results}
\label{app:cn-scheme-results}

In this appendix, we provide a detailed account of the HF results obtained in the charge-neutrality (CN) scheme, complementing the phase diagram shown in the main text.

\subsection{Numerical setup}
\label{app:cn:setup}

We study R4G for electron densities
$n_e\in[0.2,0.4]\times10^{12}\,\mathrm{cm}^{-2}$
and displacement fields
$V_{\text{disp}}\in[40,100]\,\mathrm{meV}$.
As discussed in the main text and in \appref{app:interaction-details}, in this regime the conduction electrons are concentrated near the bottom of the first conduction band of R4G in the ${\pm}K$ valleys.
Therefore, we carry out the projected HF calculations in the first conduction band of the continuum model expanded around ${\pm}K$.

At a fixed $(n_e,V_{\text{disp}})$, we discretize the continuum momentum $\bsl{k}$ with a dense triangular momentum mesh,
$\{\bsl{k}\}=\operatorname{span}_{\mathbb Z}\{\mathbf B_1,\mathbf{B}_2\}$, which extends to a maximum norm $k_{\max}$.
The same momentum mesh $\{\bsl{k}\}$ is used for all HF initial states.
We construct different WC ansatz states, with one electron per Wigner unit cell, by imposing different Wigner Bloch-momentum conservation laws throughout the HF iterations.
Each crystalline-momentum conservation rule is specified by an integer matrix $M$ that relates the primitive Wigner reciprocal lattice vectors $(\mathbf b_1,\mathbf b_2)$ to the underlying momentum mesh through
\eq{
(\mathbf b_1\ \mathbf b_2)
=
(\mathbf B_1\ \mathbf B_2)M.
}
Thus, we impose the constraint $P_{\bsl{k},\bsl{k}'}=0$ whenever ${\bsl{k}-\bsl{k}'}\notin\operatorname{span}_{\mathbb Z}\{\mathbf b_1,\mathbf b_2\}$.
For example, the $C_3$-invariant triangular lattice for $N_e=144$ corresponds to
$M=\operatorname{diag}(12,12)$ at the filling of one electron per Wigner unit cell.
Consequently, at one electron per Wigner unit cell,
$n_e = \det M \det(\mathbf{B}_1\;\mathbf{B}_2) / (2\pi)^2$, and $N_e=\det M$.
Thus, $N_e$ and $M$ determine the momentum resolution. The cutoff $k_{\max}$ is set to include all momentum states needed for reliable convergence of the HF solutions. The procedure for determining $k_{\max}$ is presented after the discussion of the HF iteration.

Using the same underlying momentum mesh for all candidate lattices is crucial for ensuring consistent finite-size effects across all WC lattice candidates.
Since the regular $\bsl{k}$ meshes of different Wigner BZs are not compatible, using a unified $\bsl{k}$ mesh is the only way to achieve consistent thermodynamic scaling and comparison among the candidate solutions.

For each candidate Wigner lattice, we solve the HF equations starting from 2--6 random initial density matrices. The number used depends on the energy variation within that Wigner lattice and on whether its lowest energy scales consistently as $k_{\max}$ or $N_e$ increases.
The HF routine iteratively updates the density matrix $P$ by diagonalizing $H_{\text{HF}}[P]$ and filling the lowest $N_e$ states, while imposing the momentum-conservation rule at every iteration.
The update of $P$ uses a %
linear mixture of $P^{(n-1)}$ and $P^{(n)}$, \ie, the density matrices from the previous and current iterations, respectively.
The convergence criterion is set to
$\|P^{(n)}-P^{(n-1)}\|_{\max}<1.5\times10^{-5}$,
where $\|P\|_{\max}\equiv\max_{ij}|P_{ij}|$ is the elementwise maximum norm.
For $N_e=144$,
this results in an energy change below
$5.3\times10^{-8}\,\text{meV}$ per electron and a self-consistency residual
$\|[H_{\text{HF}}[P],P]\|_{\max}<5\times10^{-4}\,\text{meV}$.

After obtaining convergence for all initial states, we identify the lowest-energy HF state.
To ensure a controlled comparison between $C_3$-invariant and nematic solutions, we always include at least one $C_3$-invariant Wigner lattice candidate, namely the triangular lattice with
$\tau=e^{i\pi/3}$.

We now discuss the choice of $k_{\max}$, which is set using two criteria derived from cheaper calculations at $N_e=36$. First, increasing $k_{\max}$ either (i) leaves $M$ for the minimum-energy WC unchanged up to approximate symmetries such as reflections and rotations or (ii) produces competing low-energy matrices $M$ that recur over a range of $k_{\max}$. (The competing states arise partly from approximate symmetries of the Wigner lattice.) Second, the separation of all low-energy solutions from higher-energy solutions is $10^{-4}\,\text{meV}$ per electron and persists at larger $k_{\max}$.
This results in the $k_{\max}$ values in \cref{fig:CN-phase-diagram-6x6}(n).

To reduce the computational cost, we first perform calculations for $N_e=36$, shown in \cref{fig:CN-phase-diagram-6x6}.
Here, we use both spin-valley-polarized and randomly spin-valley-distributed initial states.
All WC candidates with $M = \left(\begin{smallmatrix}
a & b\\
c & d
\end{smallmatrix}\right)
$ such that $\det M=36$, $b,c\in[-\sqrt{N_e}/2,\sqrt{N_e}/2]$, and $a,d\in[-\sqrt{N_e}-2,\sqrt{N_e}+2]$ are calculated until a convergent solution for each of them is obtained.
More anisotropic matrices $M$ have also been sampled, and none have an energy comparable to that of the identified HF ground state.
The lowest-energy Wigner lattice found at $N_e=36$, the $C_3$-invariant lattice, and nearby lattice configurations obtained through small changes to $M$ are then used as candidate lattices for the $N_e=144$ calculations that generate the phase diagram in the main text. These nearby configurations need not be multiples of any matrix $M$ used at the smaller $N_e$.
In many cases, we also perform an extended scan of WC candidates, similar to that for $N_e=36$, to ensure a sufficiently broad pool of candidates.
The computational cost at $N_e=144$ is further reduced by the fact that all HF ground states at $N_e=36$ are fully spin-valley-polarized, as seen in \cref{fig:CN-phase-diagram-6x6}(d) and (e). Therefore, we use only fully spin-valley-polarized initial states at $N_e=144$.

\subsection{Results}
\label{app:cn-result}

\begin{figure}[t]
\centering
\begin{tabular}{@{}c@{}}
\begin{tabular}{@{}l@{}}
\IfFileExists{figures/phase_diagram_app/phase_diagram_cn_6x6.pdf}{\includegraphics[width=0.99\textwidth]{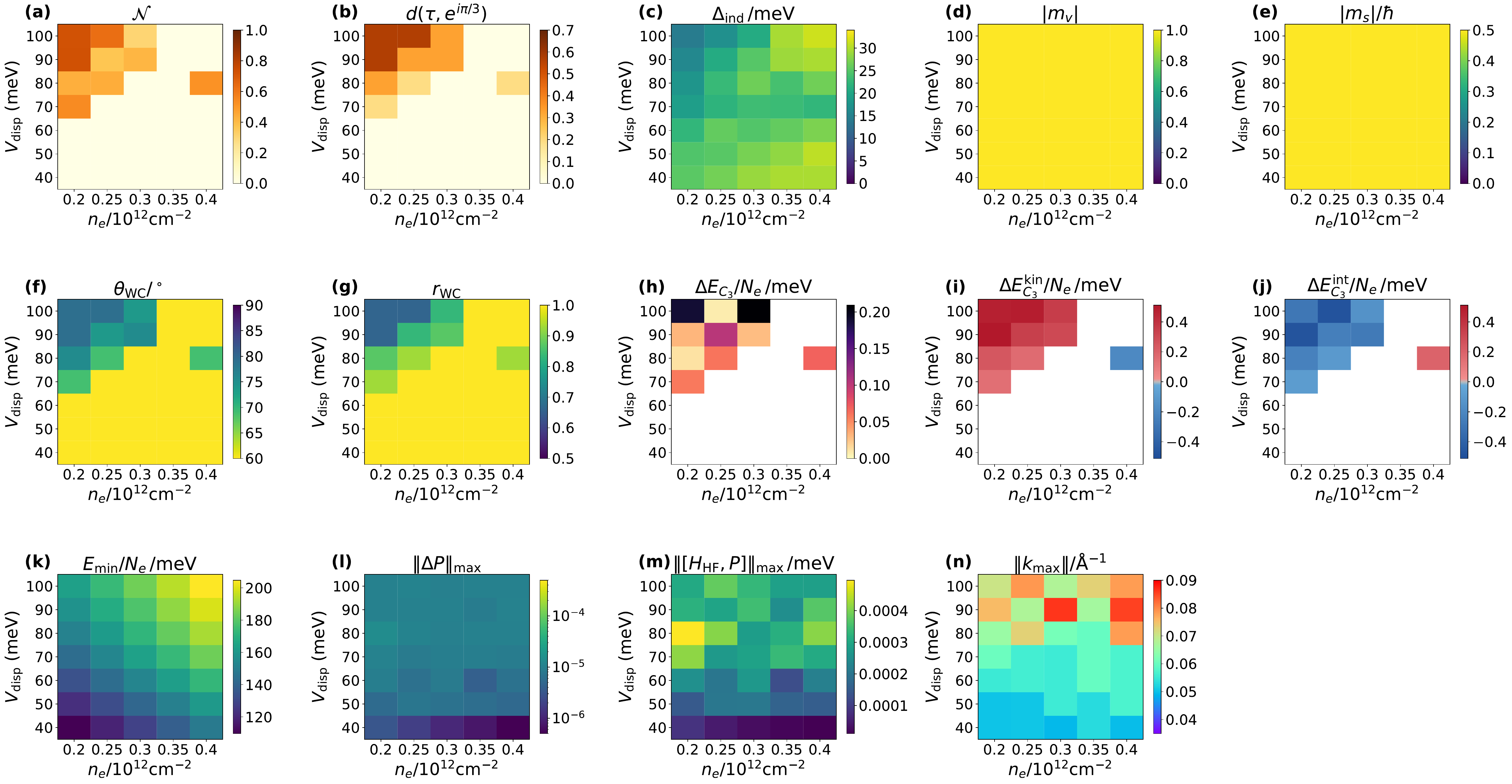}}{\fbox{\begin{minipage}[c][0.14\textheight][c]{0.46\textwidth}\centering Spin polarization of 6x6.\end{minipage}}}
\end{tabular}
\end{tabular}
\caption{
\label{fig:CN-phase-diagram-6x6}
Phase diagram of the CN-scheme results for $N_e=36$.
(a) $C_3$-symmetry violation in the density matrix, quantified by~$\mathcal{N}$.
(b)~Hyperbolic distance $d(\tau,e^{i\pi/3})$ between the Wigner lattice of the HF ground state, represented by $\tau$, and the $C_3$-invariant lattice, represented by $e^{i\pi/3}$.
(c)~HF indirect charge gap, $\Delta_\text{ind}$.
(d) Valley polarization $m_v$,
showing that the entire phase diagram is fully polarized.
(e)~Spin polarization $m_s$, showing that the entire phase diagram is fully polarized. 
(f)~Angle between the Wigner lattice basis vectors, $\theta_\text{WC}$.
For a $C_3$-invariant WC, $\theta_\text{WC}=60^\circ$.
(g)~Aspect ratio of the Wigner lattice basis vectors, $r_\text{WC}$.
For a $C_3$-invariant WC, $r_\text{WC}=1$.
(h)~Energy per electron of the lowest-energy $C_3$-invariant WC solution minus that of the HF ground state, $\Delta E_{C_3}/N_e=({E}_{C_3}-{E}_{\min})/N_e$.
(i)~Kinetic energy per electron of the lowest-energy $C_3$-invariant WC solution minus that of the HF ground state, $\Delta{E}^\text{kin}_{C_3}/N_e=({E}^\text{kin}_{C_3}-{E}^\text{kin}_{\min})/N_e$.
White regions have $C_3$-invariant HF ground states.
(j)~Interaction energy per electron of the lowest-energy $C_3$-invariant WC solution minus that of the HF ground state, $\Delta{E}^\text{int}_{C_3}/N_e=({E}^\text{int}_{C_3}-{E}^\text{int}_{\min})/N_e$.
White regions have $C_3$-invariant HF ground states.
(k)~HF ground-state energy per electron, $E_\text{min}/N_e$.
(l)~Elementwise maximal change in the density matrix in the last HF iteration, $\|\Delta{P}\|_{\max}$.
(m)~Elementwise maximum norm of the commutator, $\|[H_{\text{HF}}[P],P]\|_{\max}$.
(n)~Maximum momentum mesh cutoff, $\|k_{\max}\|$.
}
\end{figure}

\begin{figure}[t]
\IfFileExists{figures/phase_diagram_app/phase_diagram_cn.pdf}{
\includegraphics[width=0.99\textwidth]{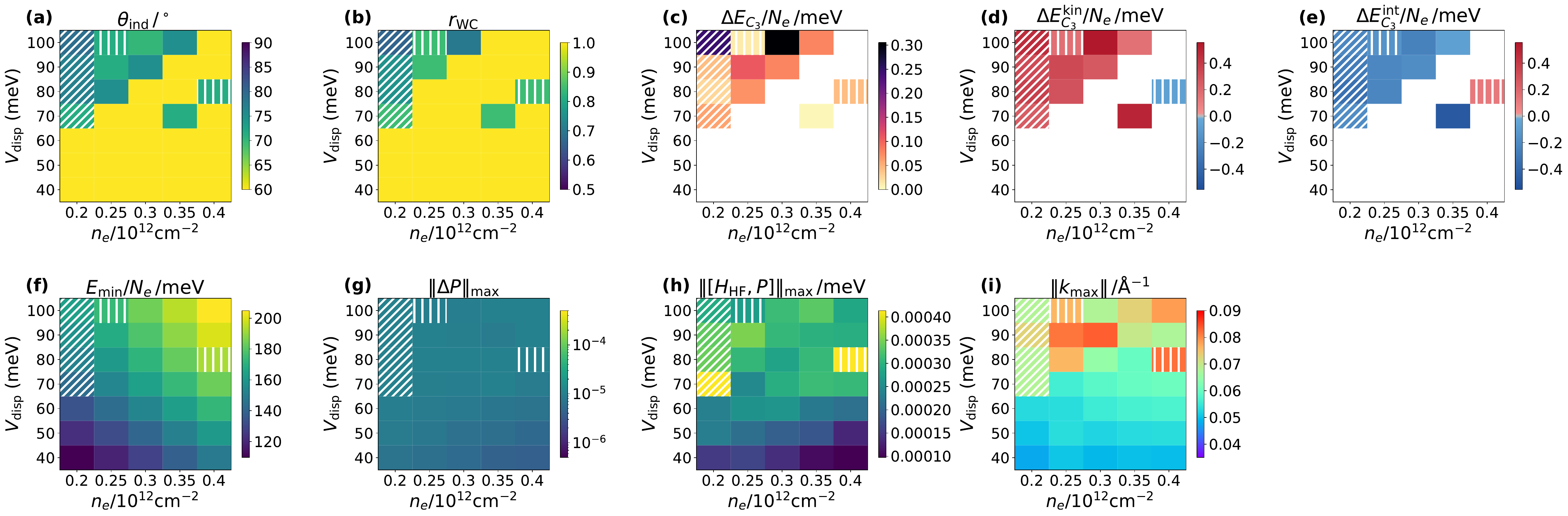}}{\fbox{\begin{minipage}[c][0.14\textheight][c]{0.46\textwidth}\centering CN scheme.\end{minipage}}}
\caption{
\label{fig:CN-phase-diagram-12x12-app}
Further details of the phase diagram of the CN-scheme results for $N_e=144$.
Vertical hatch indicates instability towards doubling of the Wigner unit cell, producing unit cells with two electrons each.
Diagonal hatch indicates instability towards metallic WCs, Fermi liquids, or stripe phases.
(a) Angle between the Wigner lattice basis vectors, $\theta_\text{WC}$.
For a $C_3$-invariant WC, $\theta_\text{WC}=60^\circ$.
(b) Aspect ratio of the Wigner lattice basis vectors, $r_\text{WC}$.
For a $C_3$-invariant WC, $r_\text{WC}=1$.
(c) Energy per electron of the lowest-energy $C_3$-invariant WC solution minus that of the HF ground state, $\Delta E_{C_3}/N_e=({E}_{C_3}-{E}_{\min})/N_e$.
(d) Kinetic energy per electron of the lowest-energy $C_3$-invariant WC solution minus that of the HF ground state, $\Delta{E}^\text{kin}_{C_3}/N_e=({E}^\text{kin}_{C_3}-{E}^\text{kin}_{\min})/N_e$.
White regions have $C_3$-invariant HF ground states.
(e) Interaction energy per electron of the lowest-energy $C_3$-invariant WC solution minus that of the HF ground state, $\Delta{E}^\text{int}_{C_3}/N_e=({E}^\text{int}_{C_3}-{E}^\text{int}_{\min})/N_e$.
White regions have $C_3$-invariant HF ground states.
(f) HF ground-state energy per electron, $E_{\min}/N_e$.
(g) Elementwise maximal change in the density matrix in the last HF iteration, $\|\Delta{P}\|_{\max}$.
(h)~Elementwise maximum norm of the commutator, $\|[H_{\text{HF}}[P],P]\|_{\max}$.
(i) Maximum momentum mesh cutoff, $\|k_{\max}\|$.
}
\end{figure}

Our computational results for $N_e=36$ are shown in \cref{fig:CN-phase-diagram-6x6}, and those for $N_e=144$ are shown in \cref{fig:cn_phase_diagram,fig:CN-phase-diagram-12x12-app}.
We first discuss the results for $N_e=36$. The HF ground states are all converged WCs, each with a sizable indirect gap $\Delta_\text{ind}>14\,\text{meV}$ throughout the phase diagram [\cref{fig:CN-phase-diagram-6x6}(c)].
Upon convergence, the maximum elementwise change in the density matrix is $\|\Delta P\|_{\max}=1.4\times10^{-5}$ [\cref{fig:CN-phase-diagram-6x6}(l)].
The maximum change in energy per electron is $1.31\times10^{-7}\,\text{meV}$,
and the maximum self-consistency residual is
$\|[H_{\text{HF}}[P],P]\|_{\max}<5\times10^{-4}\,\text{meV}$ [\cref{fig:CN-phase-diagram-6x6}(m)].

\cref{fig:CN-phase-diagram-6x6}(d) and (e) further reveal that the entire phase diagram is fully spin- and valley-polarized.
Here, the spin polarization $m_s$ and the valley polarization $m_v$ are defined by
\eq{
	|m_s| = \|\langle \vec{S}\rangle\|/N_e, \qquad |m_v| = |N_K - N_{-K}|/N_e.
}
The maximum deviation of $m_s/\hbar$ from $1/2$ is $3.4\times10^{-5}$, while the maximum deviation of $m_v$ from $1$ is $7.3\times10^{-5}$.
Full polarization is expected in R4G \cite{Zhang_2011_Chiral_FLG_QH,auerbach2025isospin,Kwan2024MFCI-III}, as it is favored by the exchange interaction.
We further note that the HF ground states throughout the phase diagram have zero Chern number.

The nematicity of the ground state is shown in Figs.~\ref{fig:CN-phase-diagram-6x6}(a) and (b) using the $C_3$ violation in the density matrix, $\mathcal{N}$, and the hyperbolic lattice distance to the $C_3$-invariant Wigner lattice $d(\tau,e^{i\pi/3})$, respectively. They have been introduced in \appref{app:hyperbolic-distance}.
The phase diagram shows that the nWCs become the ground state at lower $n_e$ and higher $V_\text{disp}$, as well as in a region of higher $n_e$ and intermediate $V_\text{disp}$.
This is further confirmed by Figs.~\ref{fig:CN-phase-diagram-6x6}(f) and (g), which show, respectively, the angle $\theta_\text{WC}$ and aspect ratio $r_\text{WC}$ of the WC primitive vectors,  $\mathbf{a}^{\text{WC}}_{1}$ and $\mathbf{a}^{\text{WC}}_{2}$.
The angle is 60${}^\circ$ for a $C_3$-invariant Wigner crystal, while $r_{\text{WC}} \equiv \min(\|\mathbf{a}^{\text{WC}}_2\|/\|\mathbf{a}^{\text{WC}}_1\|,\|\mathbf{a}^{\text{WC}}_1\|/\|\mathbf{a}^{\text{WC}}_2\|)$ is 1 in the $C_3$-invariant cases.
These quantities provide more familiar measures of nematicity in the nWC ground states.
\cref{fig:CN-phase-diagram-6x6}(h) shows that the energy per electron of each nWC solution in these regions is consistently lower than that of the lowest-energy $C_3$-invariant solution.
Furthermore, in all but one of the nWCs, the nematic ground states outcompete the corresponding $C_3$-invariant solutions through a lower kinetic energy [\cref{fig:CN-phase-diagram-6x6}(i)], at the cost of a higher interaction energy [\cref{fig:CN-phase-diagram-6x6}(j)]. This agrees with the results for the larger system with $N_e=144$, discussed below. The outlying case, at  higher $n_e$ and intermediate $V_\text{disp}$, is found to be unstable according to the TDHF calculation for $N_e=144$.

Now we move on to the phase diagram for $N_e=144$ (\cref{fig:cn_phase_diagram,fig:CN-phase-diagram-12x12-app}).
Here, only fully spin-valley-polarized initial states are used.
We also use $k_{\max}$ values determined from the $N_e=36$ calculations [\cref{fig:CN-phase-diagram-6x6}(n)] to set $k_{\max}$ for $N_e=144$ [\cref{fig:CN-phase-diagram-12x12-app}(i)].
Upon convergence, the maximum elementwise change in the density matrix is $1.02\times10^{-5}$ [\cref{fig:CN-phase-diagram-12x12-app}(g)], the energy change is below
$5.3\times10^{-8}\,\text{meV}$ per electron, and the self-consistency residual satisfies
$\|[H_{\text{HF}}[P],P]\|_{\max}<5\times10^{-4}\,\text{meV}$ [\cref{fig:CN-phase-diagram-12x12-app}(h)].

The nematicity in the $N_e=144$ calculations is discussed in the main text. Here, we present the WC lattice geometry in detail: the angle between the primitive lattice vectors $\theta_\text{WC}$ in \cref{fig:CN-phase-diagram-12x12-app}(a), and their aspect ratio $r_\text{WC}$ in \cref{fig:CN-phase-diagram-12x12-app}(b).
\cref{fig:CN-phase-diagram-12x12-app}(c) shows that, in the nWC phase, the energy of the HF ground state is consistently lower than that of the lowest-energy $C_3$-invariant solution.
As for $N_e=36$, the entire $N_e=144$ phase diagram has a Chern number of zero.

Furthermore, we present details of the energetics.
As discussed in the main text, in the CN scheme the nWC can be favored by a kinetic-energy mechanism: occupying a single trigonal-warping lobe to accommodate the Wigner reciprocal primitive vectors avoids the cost of spreading weight over higher-energy regions of all three lobes. This hypothesis has been tested directly using the converged HF solutions.
For a given HF density matrix $P$, the kinetic energy per electron is
$E^{\text{kin}} =  \sum_{\bsl{k}\eta s} \varepsilon_{\eta}(\bsl{k})\, \langle \gamma^\dagger_{\bsl{k}\eta s} \gamma_{\bsl{k} \eta s} \rangle/N_e$,
where $\varepsilon_{\eta}(\bsl{k})$ is the energy of the lowest conduction band in the valley $\eta$.
We compare the total, kinetic, and interaction energies of the nematic HF solution with those of the lowest-energy $C_3$-invariant state at the same $(n_e,V_\text{disp})$,
reported as $\Delta E_{C_3} \equiv E_{C_3}-E_{\min}$, $\Delta E^\text{kin}_{C_3} \equiv E^\text{kin}_{C_3}-E^\text{kin}_{\min}$, and $\Delta E^\text{int}_{C_3} \equiv E^\text{int}_{C_3}-E^\text{int}_{\min}$, in Figs.~\ref{fig:CN-phase-diagram-12x12-app}(c), (d), and (e), respectively, for $N_e=144$.
In the stable nematic regions, the nematic HF ground state has lower kinetic energy [\cref{fig:CN-phase-diagram-12x12-app}(d)] and higher interaction energy [\cref{fig:CN-phase-diagram-12x12-app}(e)] than the lowest-energy $C_3$-invariant solution.
The net effect is a lower total energy for the nWC solutions.

\section{Average scheme results}
\label{app:average-scheme-result}

\begin{figure}
\IfFileExists{figures/phase_diagram_app/phase_diagram_avg.pdf}{\includegraphics[width=0.99\textwidth]{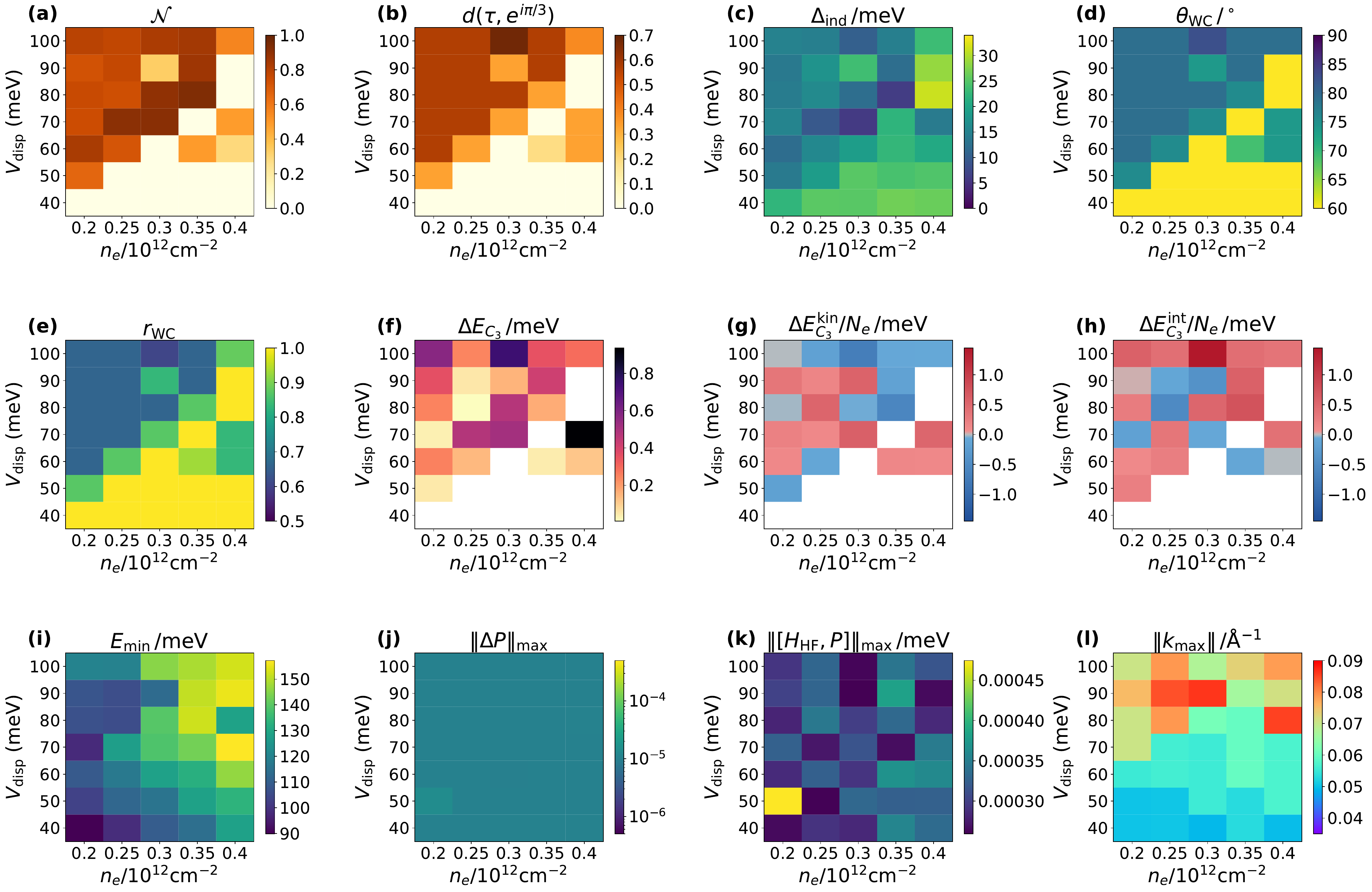}}{\fbox{\begin{minipage}[c][0.14\textheight][c]{0.46\textwidth}\centering Average scheme.\end{minipage}}}
\caption{
Phase diagram of the average-scheme results for $N_e=36$. All HF solutions are fully spin- and valley-polarized WCs.
(a) $C_3$-symmetry violation in the density matrix, quantified by $\mathcal{N}$.
(b) Hyperbolic distance to the $C_3$-invariant lattice, $d(\tau,e^{i\pi/3})$.
(c) HF indirect charge gap, $\Delta_{\text{ind}}$.
(d) Angle between the Wigner lattice basis vectors, $\theta_\text{WC}$.
(e) Aspect ratio of the Wigner lattice basis vectors, $r_\text{WC}$.
(f) Energy per electron of the lowest-energy $C_3$-invariant WC solution minus that of the HF ground state, $\Delta E_{C_3}/N_e=({E}_{C_3}-{E}_{\min})/N_e$.
(g) Kinetic energy per electron of the lowest-energy $C_3$-invariant WC solution minus that of the HF ground state, $\Delta{E}^\text{kin}_{C_3}/N_e=({E}^\text{kin}_{C_3}-{E}^\text{kin}_{\min})/N_e$.
White regions have $C_3$-invariant HF ground states.
(h) Interaction energy per electron of the lowest-energy $C_3$-invariant WC solution minus that of the HF ground state, $\Delta{E}^\text{int}_{C_3}/N_e=({E}^\text{int}_{C_3}-{E}^\text{int}_{\min})/N_e$.
White regions have $C_3$-invariant HF ground states.
(i) HF ground-state energy per electron, $E_{\text{min}}/N_e$.
(j) Elementwise maximal change in the density matrix during the last HF iteration, $\|\Delta{P}\|_{\max}$.
(k)~Elementwise maximum norm of the commutator, $\|[H_{\text{HF}}[P],P]\|_{\max}$.
(l) Maximum momentum-mesh cutoff, $\|k_{\max}\|$.
}
\label{fig:phase-avg}
\end{figure}

The average scheme differs from the CN scheme only by the reference one-body potential $H_1$ in \cref{eq:H_proj}, which is discussed in detail in \appref{app:interaction-details}.
An additional exchange-field potential due to the valence-band electrons renormalizes the single-particle dispersion used in the projected HF problem and could potentially change the physics.
Hence, we calculate the phase diagram in the average scheme for $N_e=36$, using the same initial-state sampling strategy and convergence criteria as in the CN scheme.
The result is shown in \cref{fig:phase-avg}, which matches the general features of the CN scheme.
In summary, we obtain WC ground states with an indirect gap throughout the phase diagram. The nWC phase appears at larger $V_{\text{disp}}$, both in the upper left of the phase diagram, where $V_\text{disp}$ is high and $n_e$ is low, and around $n_e\approx0.4\times10^{12}\,\text{cm}^{-2}$ and $V_\text{disp}\approx70\,\text{meV}$. This is shown by the maximum violation of $C_3$ symmetry in the density matrix, $\mathcal{N}$, in \cref{fig:phase-avg}(a) and by the hyperbolic lattice distance in \cref{fig:phase-avg}(b).
The average scheme phase diagram shows a more extended nWC region compared to that of the CN scheme.
The agreement between the two schemes confirms the emergence of an nWC in rhombohedral graphene.
Below, we discuss the average-scheme results in more detail.

We use fully spin-valley-polarized initial states for the average-scheme calculations. The maximum momentum cutoffs [\cref{fig:phase-avg}(l)] are similar to those used in the CN scheme.
Under the same convergence criterion, the maximum change in the density matrix during the last HF iteration is $\|\Delta P\|_{\max}=1.41\times 10^{-5}$ [\cref{fig:phase-avg}(j)], while the maximum change in the energy per electron is $1.76\times10^{-7}\,\text{meV}$.
All HF ground states are WCs with an indirect gap $\Delta_{\text{ind}}>5.8\,\text{meV}$ [\cref{fig:phase-avg}(c)].
Figures \ref{fig:phase-avg}(d) and (e) show the angle and aspect ratio between the Wigner lattice primitive vectors, providing more familiar descriptors of nematicity.
\cref{fig:phase-avg}(f) further shows that, in the nWC phase, the lowest-energy $C_3$-invariant solutions always have higher energies than the HF ground state.
However, compared with the CN scheme, the  energetic picture in the average scheme is more complicated.
The nWC ground states outcompete $C_3$-invariant WCs through a lower kinetic energy, or a lower interaction energy, or both, as seen in Figs.~\ref{fig:phase-avg}(g) and (h).
Finally, we note that all HF ground states have a Chern number of zero.

\section{Time-dependent Hartree--Fock}
\label{app:tdhf-details}

\subsection{Review of TDHF}

The small-amplitude limit of time-dependent {Hartree--Fock} (TDHF) about a stationary HF Slater determinant is equivalent to the self-consistent random-phase approximation (RPA), provided that the residual kernel is obtained by linearizing the same time-local HF mean field used to determine the stationary state. It approximates neutral, particle-number-conserving collective excitations, while the associated HF orbital Hessian tests the local stability of the determinant to quadratic order within the retained particle--hole variational space~\cite{Thouless1960,Thouless1961,Rowe1970NuclearCollectiveMotion,RingSchuck1980NuclearManyBody,BlaizotRipka1986FiniteSystems}.
The input to a TDHF calculation is a stationary HF one-body density matrix $P_0$ satisfying
\eq{
  [H_{\mathrm{HF}}[P_0],P_0]=0 .
  \label{eq:app:stationary-hf-density}
}
For a zero-temperature Slater determinant, $P_0=P_0^\dagger=P_0^2$. We set $\hbar=1$ throughout.
After diagonalizing $H_{\mathrm{HF}}[P_0]$, we denote its eigenstates by $|\phi_a\rangle$ and its eigenvalues by $\epsilon_a$. Occupied HF orbitals are labeled by $h,h'$, and unoccupied HF orbitals are labeled by $p,p'$.
The linearized idempotency condition,
$P_0\delta P+\delta P P_0=\delta P$, removes the occupied--occupied and unoccupied--unoccupied blocks. A complexified tangent fluctuation can therefore be written as
\eq{
  \delta P(t)
  =
  \sum_{ph}
  \left[
    X_{ph}(t)|\phi_p\rangle\langle\phi_h|
    +
    Y_{ph}(t)|\phi_h\rangle\langle\phi_p|
  \right].
  \label{eq:app:tangent-density}
}
Here $X_{ph}(t)$ and $Y_{ph}(t)$ are independent coordinates in the complexified particle--hole tangent space. Hermiticity will be imposed on the physical time-dependent fluctuation below.
Linearizing the TDHF equation $i\partial_tP=[H_{\mathrm{HF}}[P],P]$ about $P_0$ gives
\eq{
  i\partial_t\delta P
  =
  [H_{\mathrm{HF}}[P_0],\delta P]
  +
  [\delta H_{\mathrm{HF}},P_0],
  \qquad
  \delta H_{\mathrm{HF}}
  =
  \left.\frac{\delta H_{\mathrm{HF}}}{\delta P}\right|_{P_0}\!\delta P .
  \label{eq:app:linearized-tdhf}
}
Here the derivative denotes the linear Fr\'echet derivative of the HF one-body Hamiltonian with respect to the one-body density matrix.
For a normal mode $\nu$, take
$X_{ph}(t)=X_{ph}^{\nu}e^{-i\Omega_\nu t}$ and
$Y_{ph}(t)=Y_{ph}^{\nu}e^{-i\Omega_\nu t}$, and define the coefficient of $e^{-i\Omega_\nu t}$ by
\eq{
  \delta P_\nu
  =
  \sum_{ph}
  \left(
    X_{ph}^{\nu}|\phi_p\rangle\langle\phi_h|
    +
    Y_{ph}^{\nu}|\phi_h\rangle\langle\phi_p|
  \right).
  \label{eq:app:neutral-excitation}
}
The Hermitian physical density fluctuation associated with this complex mode is
\eq{
  \delta P_{\mathrm{phys},\nu}(t)
  =
  \delta P_\nu e^{-i\Omega_\nu t}
  +
  \delta P_\nu^\dagger e^{+i\Omega_\nu^*t}.
  \label{eq:app:hermitian-density-mode}
}
Its $ph$ block is
$X^\nu e^{-i\Omega_\nu t}+(Y^\nu)^*e^{+i\Omega_\nu^*t}$, and its $hp$ block is the Hermitian conjugate. Thus $X^\nu$ and $Y^\nu$ in one complex normal-mode component are not required to be mutual conjugates~\cite{RingSchuck1980NuclearManyBody,BlaizotRipka1986FiniteSystems,Kwan2024MFCI-III}.

Substitution of the harmonic ansatz into \cref{eq:app:linearized-tdhf} gives the RPA/TDHF eigenvalue problem~\cite{Thouless1961,Rowe1970NuclearCollectiveMotion,RingSchuck1980NuclearManyBody,BlaizotRipka1986FiniteSystems,Kwan2024MFCI-III}
\eq{
  \mathcal L
  \begin{pmatrix}
    X\\
    Y
  \end{pmatrix}
  =
  \Omega
  \begin{pmatrix}
    X\\
    Y
  \end{pmatrix},
  \qquad
  \mathcal L
  \equiv
  \begin{pmatrix}
    A & B\\
    -B^* & -A^*
  \end{pmatrix}.
  \label{eq:tdhf-dynamical-matrix}
}
The matrix $\mathcal L$ is the RPA dynamical matrix. Its blocks are
\eq{
  A_{ph,p'h'}
  =
  (\epsilon_p-\epsilon_h)\delta_{pp'}\delta_{hh'}
  +K^A_{ph,p'h'},
  \qquad
  B_{ph,p'h'}=K^B_{ph,p'h'}.
  \label{eq:tdhf-A-B-blocks}
}
The diagonal term $\epsilon_p-\epsilon_h$ is the unperturbed HF particle--hole energy. The kernels $K^A$ and $K^B$ describe the induced change of the Hartree and Fock fields caused by the density fluctuation.
For a Hermitian, density-independent two-body Hamiltonian, $A=A^\dagger$ and $B=B^T$. With
$\Sigma_z=\operatorname{diag}(I,-I)$ and
$\Sigma_x=\begin{pmatrix}0&I\\I&0\end{pmatrix}$, these identities imply
\eq{
  \mathcal L^\dagger\Sigma_z=\Sigma_z\mathcal L,
  \qquad
  \mathcal L=-\Sigma_x\mathcal L^*\Sigma_x.
  \label{eq:app:rpa-symmetries}
}
The first relation is pseudo-Hermiticity with respect to the indefinite RPA metric; the second produces the conjugate-pair symmetry of the complete doubled particle--hole problem. A restricted momentum or internal-charge block need not be closed under the second operation~\cite{Nakada2016,Kwan2024MFCI-III}.

To give explicit formulas for the kernels, we introduce a composite band-momentum index $i=(\bsl k,m,\eta,s)$ and write the normal-ordered projected interaction in the band basis as
\eq{
  :H_{\mathrm{int}}:
  =
  \frac{1}{2}
  \sum_{ijkl}
  U_{ij,kl}
  \gamma_i^\dagger
  \gamma_j^\dagger
  \gamma_l
  \gamma_k.
  \label{eq:projected-two-body-U}
}
Here $U_{ij,kl}$ is a non-antisymmetrized two-body matrix element with
$U_{ij,kl}=U_{ji,lk}=U_{kl,ij}^*$; the exchange terms are displayed explicitly below.
If $\hat f_\alpha^\dagger=\sum_i\psi_{i\alpha}\gamma_i^\dagger$ creates a self-consistent HF orbital $\alpha$, the interaction in the HF orbital basis is
\eq{
  :H_{\mathrm{int}}:
  =
  \frac{1}{2}
  \sum_{\alpha\beta\mu\delta}
  \mathcal U_{\alpha\beta,\mu\delta}
  \hat f_\alpha^\dagger
  \hat f_\beta^\dagger
  \hat f_\delta
  \hat f_\mu,
  \qquad
  \mathcal U_{\alpha\beta,\mu\delta}
  =
  \sum_{ijkl}
  \psi^*_{i\alpha}
  \psi^*_{j\beta}
  U_{ij,kl}
  \psi_{k\mu}
  \psi_{l\delta}.
  \label{eq:hf-basis-interaction}
}
With particle--hole labels $ph$ and $p'h'$, the TDHF kernels are
\eq{
  K^A_{ph,p'h'}
  =
  \mathcal U_{p h',h p'}
  -
  \mathcal U_{p h',p' h},
  \qquad
  K^B_{ph,p'h'}
  =
  \mathcal U_{p p',h h'}
  -
  \mathcal U_{p p',h' h}.
  \label{eq:tdhf-kernels}
}
The first term in each expression is the direct contribution and the second is the exchange contribution. These formulas use the same convention as the stationary HF Hamiltonian~\cite{Kwan2024MFCI-III}.
These expressions assume a fixed, density-independent two-body interaction. A density-dependent effective interaction or energy functional would add rearrangement terms through
$\delta H_{\mathrm{HF}}/\delta P$. Such term does not appear in the fixed Coulomb interaction for this work.

The same interaction functional generates both the linearized TDHF matrix and the HF orbital Hessian. The Hessian, rather than the eigenvalues of the non-Hermitian dynamical matrix alone, is the direct variational stability test.
The eigenvectors of $\mathcal L$ carry the indefinite RPA norm
\eq{
  \mathcal N_\nu
  =
  Z_\nu^\dagger\Sigma_z Z_\nu
  =
  X_\nu^\dagger X_\nu
  -
  Y_\nu^\dagger Y_\nu,
  \qquad
  Z_\nu=
  \begin{pmatrix}
    X_\nu\\
    Y_\nu
  \end{pmatrix}.
  \label{eq:rpa-indefinite-norm}
}
By \cref{eq:app:rpa-symmetries}, if $Z_\nu$ has eigenvalue $\Omega_\nu$, then $\Sigma_xZ_\nu^*$ has eigenvalue $-\Omega_\nu^*$ and the opposite RPA norm. If the stability matrix defined below is positive semidefinite, every nonzero-frequency solution belongs to a real-frequency physical pair. The creation branch may be chosen with $\Omega_\nu>0$ and $\mathcal N_\nu>0$ and normalized to $\mathcal N_\nu=1$; its negative-frequency, negative-norm partner is the corresponding de-excitation~\cite{Nakada2016,Nakada2016add,Kwan2024MFCI-III}.
Accordingly, for a stable solution we retain the positive-frequency, positive-norm branch when quoting finite-frequency collective modes. An isolated zero-norm eigenvector at nonzero frequency cannot be normalized as a quasiboson excitation.
Zero-frequency solutions require a more delicate analysis \cite{Cui2013,Nakada2016,Nakada2016add}.

As a numerical sanity check, the full unfiltered spectrum must exhibit the RPA pairing $\Omega,-\Omega^*$, provided that the retained particle--hole space includes both members of every particle--hole-conjugate pair.
This check is a necessary but not sufficient condition for stability or normalizability.

The Hermitian RPA stability matrix is
\eq{
  \mathcal S
  \equiv
  \Sigma_z\mathcal L
  =
  \begin{pmatrix}
    A & B\\
    B^* & A^*
  \end{pmatrix}.
  \label{eq:tdhf-stability-matrix}
}
When $\mathcal S$ is indefinite, RPA may contain imaginary or genuinely complex frequencies, as well as positive-frequency solutions with negative or vanishing norm; zero eigenvalues may additionally require Jordan-chain analysis~\cite{Nakada2016,Cui2013}.
Thus, $\mathcal{S}$ encodes the local stability of HF solutions.

To state the variational meaning of $\mathcal S$, parameterize a nearby Slater determinant by the anti-Hermitian occupied--unoccupied orbital rotation
\eq{
  \hat\kappa =
  \sum_{ph}
  \left(
    \kappa_{ph}\hat f_p^\dagger\hat f_h
    -
    \kappa_{ph}^*\hat f_h^\dagger\hat f_p
  \right),
  \qquad
  |\Phi(\kappa)\rangle=e^{\hat\kappa}|\Phi_0\rangle .
  \label{eq:app:orbital-rotation}
}
With the doubled physical tangent vector
$\mathcal K=(\kappa\;\kappa^*)^T$, stationarity removes the linear term and, with the present normalization convention,
\eq{
  E[\kappa]-E[0]
  =
  \frac{1}{2}\mathcal K^\dagger\mathcal S\mathcal K
  +O(\lVert\kappa\rVert^3).
  \label{eq:app:hf-second-variation}
}
The particle--hole conjugation symmetry of $\mathcal S$ makes each of its eigenspaces invariant under $\mathcal K\mapsto\Sigma_x\mathcal K^*$; consequently, its curvature can be tested on physical fixed vectors of the form $\mathcal K=(\kappa\;\kappa^*)^T$. Thus, $\mathcal S>0$ gives a strict local HF minimum within the retained active-space Slater-determinant manifold. The weaker condition $\mathcal S\geq0$ establishes only non-negative quadratic curvature: symmetry directions may be flat, while any other zero direction requires higher-order analysis. A zero eigenvalue is a Goldstone direction only when its vector is generated by an exact continuous symmetry that the HF state breaks~\cite{Thouless1960,Cui2013}.
A negative eigenvalue of $\mathcal S$ provides a physical particle--hole orbital-rotation direction along which the HF energy decreases and therefore proves the local quadratic instability of the assumed WC state.
Note that again $\mathcal{S}$ encodes only necessary but not sufficient conditions for establishing the HF solution as a global minimum.

The TDHF computation can be simplified using momentum conservation. At fixed conserved crystal momentum $\bsl q$, defined modulo a WC reciprocal vector, define
\eq{
  \hat Q_{\nu,\bsl q}^\dagger
  =
  \sum_{ph,\bsl k}
  \left[
    X^\nu_{ph}(\bsl k,\bsl q)
    \hat f^\dagger_{p,\bsl k+\bsl q}
    \hat f_{h,\bsl k}
    -
    Y^\nu_{ph}(\bsl k,\bsl q)
    \hat f^\dagger_{h,\bsl k}
    \hat f_{p,\bsl k-\bsl q}
  \right].
  \label{eq:app:momentum-rpa-operator}
}
Both operator terms carry total crystal momentum $\bsl q$. The $X$ amplitudes are indexed by underlying particle--hole transitions with momentum $+\bsl q$, whereas the $Y$ amplitudes are indexed by the reversed transitions with momentum $-\bsl q$.
Equation~\eqref{eq:tdhf-dynamical-matrix} is then diagonalized separately in each $\bsl q$ sector.
Softening of a positive-norm mode to zero at nonzero $\bsl q$ marks marginality and can signal the onset of competing neutral order at that wave vector. A local quadratic instability is established directly by a negative eigenvalue of the corresponding Hermitian Hessian block $\mathcal S(\bsl q)$, which combines the $+\bsl q$ and $-\bsl q$ tangent amplitudes as in \cref{eq:app:momentum-rpa-operator}~\cite{Thouless1960,Nakada2016,Cui2013}.
If the HF state spontaneously breaks an exact continuous symmetry, the corresponding infinitesimal generator gives a zero-curvature direction provided that the active-space truncation and numerical implementation preserve the symmetry. Its momentum is fixed by how the broken generator transforms under the unbroken translations; it is $\bsl q=0$ for uniform internal rotations and for continuous translations represented in the WC reduced zone. Explicit lattice pinning, spin anisotropy, a symmetry-violating projection, or numerical symmetry breaking can gap the associated collective mode. Accidental or critical zero modes can also occur without a broken continuous symmetry~\cite{Cui2013,Kwan2024MFCI-III}.
Finally, the conjugate-pair operation maps a momentum-$\bsl q$ eigenvector to a momentum-$-\bsl q$ eigenvector and reverses any additive internal charge carried by the mode. Hence, if $\Omega_\nu(\bsl q)$ occurs in one block, then $-\Omega_\nu(\bsl q)^*$ occurs in the particle--hole-conjugate block at $-\bsl q$. The $\Omega,-\Omega^*$ check therefore applies only after combining the $\bsl q$ and $-\bsl q$ spectra and, when applicable, the opposite internal-charge sectors; it need not appear within one generic block~\cite{Kwan2024MFCI-III}.

\subsection{Results}

\begin{figure}
	\centering
    \begin{minipage}{\columnwidth}
		\centering
		\begin{tabular}{@{}c@{}}
			\begin{tabular}{@{}l@{}}\IfFileExists{figures/tdhf_cn/TDHF_pd_dispersion.pdf}{\includegraphics[width=0.98\textwidth]{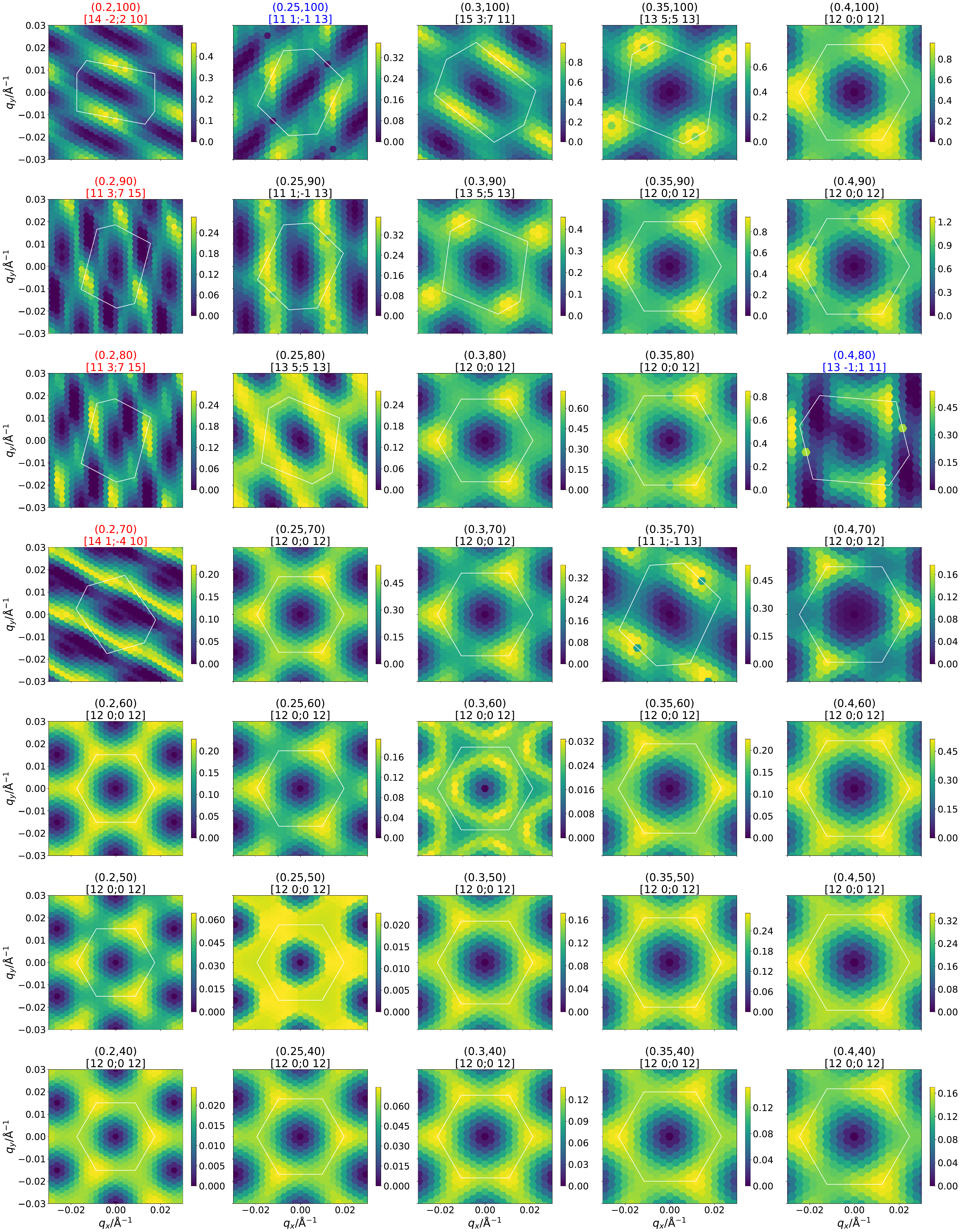}}{\fbox{\begin{minipage}[c][0.14\textheight][c]{0.46\textwidth}\centering TDHF neutral mode spectra across phase diagram\end{minipage}}}\end{tabular}
		\end{tabular}
	\end{minipage}
	\caption{
    Lowest-energy TDHF neutral-mode dispersion in meV at each $(n_e,V_{\text{disp}})$ across the phase diagram, labeled by the Wigner lattice matrix $M$.
    Here, $n_e$ is in units of $10^{12}\,\text{cm}^{-2}$ and $V_{\text{disp}}$ is in meV.
    Blue labels indicate instability towards Wigner unit cell doubling that can lead to two electrons per cell.
    Red labels indicate instability towards metallic WCs, Fermi liquids, or stripe phases.
    White lines show the boundary of the first Wigner Brillouin zone.}
	\label{fig:TDHF-dispersion-phase-diagram}
\end{figure}

\begin{figure}
	\centering
	\begin{minipage}{\columnwidth}
		\centering
		\begin{tabular}{@{}c@{}}
			\begin{tabular}{@{}l@{}}\IfFileExists{figures/tdhf_cn/TDHF_pd_stability.pdf}{\includegraphics[width=0.98\textwidth]{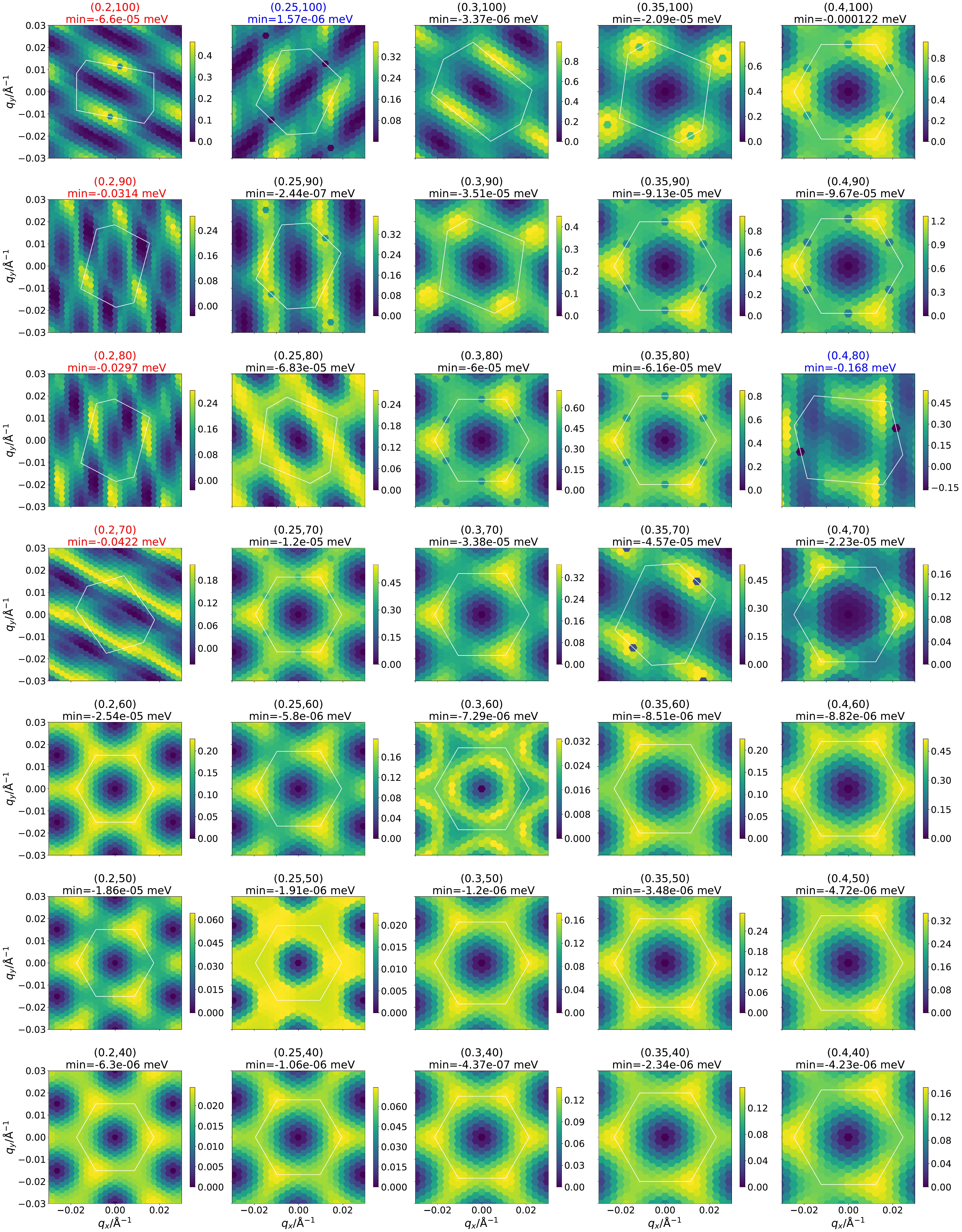}}{\fbox{\begin{minipage}[c][0.14\textheight][c]{0.46\textwidth}\centering TDHF stability spectra across phase diagram\end{minipage}}}\end{tabular}
		\end{tabular}
	\end{minipage}
	\caption{Smallest TDHF stability eigenvalue in meV at each $(n_e,V_{\text{disp}})$ across the phase diagram, with the minimum stability eigenvalue labeled. Here, $n_e$ is in units of $10^{12}\,\text{cm}^{-2}$ and $V_{\text{disp}}$ is in meV.
    Blue labels indicate instability towards Wigner unit cell doubling that can lead to two electrons per cell.
    Red labels indicate instability towards metallic WCs, Fermi liquids, or stripe phases.
    White lines show the boundary of the first Wigner Brillouin zone.}
	\label{fig:TDHF-stability-phase-diagram}
\end{figure}

In \cref{fig:TDHF-dispersion-phase-diagram,fig:TDHF-stability-phase-diagram},
we present TDHF diagnostics across the phase diagram.
Figure~\ref{fig:TDHF-dispersion-phase-diagram} shows the lowest positive-norm
neutral-mode energy in each momentum sector, while
\cref{fig:TDHF-stability-phase-diagram} shows the corresponding smallest
eigenvalue of the stability matrix $\mathcal{S}$.
In general, negative eigenvalues of $\mathcal{S}$ indicate instability of the HF solution.
Due to numerical resolution, we adopt the following criterion for the stability of the HF solution in our case.
Let $\xi(\bsl{q})$ be the stability eigenvalue of a HF solution at momentum $\bsl{q}$.
A solution is deemed stable if $\xi(\mathbf{0}) > 5 \|[H_\text{HF},P]\|_{\max}$, where $\|\cdot\|_{\max}$ denotes the elementwise maximum norm, and if
$\xi(\bsl{q}) > 10^{-4}\,\mathrm{meV}$
at every local minimum of $\xi(\bsl{q})$ with $\bsl{q}\ne\mathbf{0}$.

In the $C_3$-invariant WC phase, all computed nonzero positive-norm modes have positive energies, while the expected Goldstone modes occur at zero energy at $\bsl q=0$.
This is consistent, within our numerical criterion, with the local stability of the $C_3$-invariant WC, as further shown by the lowest stability eigenvalue labeled in each panel of \cref{fig:TDHF-stability-phase-diagram}. At $\bsl{q}\ne0$, all the lowest stability eigenvalues are positive, while at $\bsl{q}=0$ they are zero within $5 \|[H_\text{HF},P]\|_{\max}$.

For nWCs, the behavior differs between stable and unstable solutions. Under the same stability criterion, the stable cases are similar to the $C_3$-invariant WC, except that their spectra are no longer $C_3$ symmetric.
For the unstable cases, all of them have local minima at nonzero $\bsl{q}$, signaling the existence of soft modes.
There are two kinds of instability.
The first kind is labeled in blue in \cref{fig:TDHF-stability-phase-diagram}.
Their local minima are located at half of a Wigner primitive reciprocal lattice vector, which can result in doubling of the Wigner unit cell. This can lead to a WC with two electrons per cell.
These cases occur at high $V_\text{disp}$ and low $n_e$, as well as at intermediate $V_\text{disp}$ and higher $n_e$.
The second kind of instability is labeled in red in \cref{fig:TDHF-stability-phase-diagram}.
Their local minima are located at generic $\bsl{q}$, which can result in incommensurate WC fillings and give rise to either a metallic WC~\cite{Han_2026_Wigner_R4G,Dong_2026_mWC_RMG} or other phases, including Fermi liquids and stripe phases.
These cases occur at high $V_\text{disp}$ and the lowest $n_e$ values.
We note that, although soft modes at $\mathbf{b}/2$ appear in many regions of the $C_3$-invariant WC and nWC phases, in most cases they do not develop into instabilities.

\end{document}